\documentclass[12pt]{iopart}

\usepackage{graphicx}

\usepackage{float}
\usepackage{placeins}
\usepackage[super]{nth}

\usepackage{upgreek}

\begin{document}

\title[Parametric Waveform Synthesis]{Parametric Waveform Synthesis: a scalable approach to generate sub-cycle optical transients}

\author{Roland E. Mainz$^{1,2,\dagger}$, Giulio Maria Rossi$^{1,2,\dagger}$, Fabian Scheiba$^{1,2}$, Miguel A. Silva-Toledo$^{1,2}$, Yudong Yang$^{1,2}$, Giovanni Cirmi$^{1,2}$, and Franz X. K\"artner$^{1,2}$}

\address{$^1$Center for Free-Electron Laser Science, Deutsches Elektronen-Synchrotron DESY, Notkestraße 85, 22607 Hamburg, Germany\\
$^2$Physics Department and The Hamburg Centre for Ultrafast Imaging, University of Hamburg, Luruper Chaussee 149, 22761 Hamburg, Germany\\
$\dagger$ These authors contributed equally}

\ead{roland.mainz@desy.de}
\vspace{10pt}
\begin{indented}
\item[]October 2022
\end{indented}

\begin{abstract}
The availability of electromagnetic pulses with controllable field waveform and extremely short duration, even below a single optical cycle, is imperative to fully harness strong-field processes and to gain insight into ultrafast light-driven mechanisms occurring in the attosecond time-domain. The recently demonstrated parametric waveform synthesis (PWS) introduces an energy-, power- and spectrum-scalable method to generate non-sinusoidal sub-cycle optical waveforms by coherently combining different phase-stable pulses attained via optical parametric amplifiers. Significant technological developments have been addressed to overcome the stability issues related to PWS and to obtain an effective and reliable waveform control system. Here we present the main ingredients enabling PWS technology. The design choices concerning the optical, mechanical and electronic setups are justified by analytical/numerical modeling and benchmarked by experimental observations. In its present incarnation, the PWS technology enables the generation of field-controllable mJ-level few-femtosecond pulses spanning the visible to infrared range. 
\end{abstract} 


%
%
%
%
%

\clearpage

\section{Introduction to Parametric Waveform Synthesis}
Over the past three decades, the development of ultrabroadband optical parametric amplifiers (OPAs) started a new era for ultrashort pulse generation\cite{Cerullo_OptLett_1998,Cerullo_RevSciInstr_2003,Brida_JOpt_2010}. Together with spectral broadening and compression techniques, OPAs allowed generating high-energy pulses with durations significantly shorter than the pump pulse from which they originate. Originally developed with sub-ps duration pump lasers, the OPA has been adapted to longer pump pulses. In this case, it is often called optical parametric chirped-pulse amplification (OPCPA). Most of the analysis presented in this article applies to both OPA and OPCPA, in this case we will refer to it as OP(CP)A.\\
Nonlinear broadening/compression techniques such as hollow-core fiber compressors \cite{Nisoli_AptPhysLett_1996} (HCFC), multi-plate continua \cite{Lu2014} (MPC) or Heriott cells \cite{Schulte2016}, allow broadening the spectrum of the initial pulse in an almost symmetrical way, therefore without strongly altering its central wavelength. These techniques have the advantage of being very efficient, but they usually do not allow to attain high-energy pulses spanning significantly more than one octave of bandwidth. Therefore, standard broadening/compression techniques do not allow for the generation of high-energy pulses with durations below a single optical cycle, that requires significantly more than one octave of bandwidth. Multi-octave spanning spectra are also necessary to support non-sinusoidal waveforms. Sub-cycle non-sinusoidal IR waveforms are, for instance, applied in strong-field interactions such as high-harmonic generation (HHG), allowing for the generation of isolated attosecond pulses \cite{Rossi_NatPhot_2020} and their tunability over the XUV/soft X-ray range \cite{Yang_NatComm_2021}.\\
The OP(CP)A process enables to tune the amplification bandwidth over a wide spectral region \cite{Cerullo_OptLett_1998,Brida_JOpt_2010}. By exploiting different materials and optical schemes, the OPA bandwidth can be almost continuously tuned across a large portion of the optical spectrum, approximately spanning the visible to mid-IR region. Moreover, OP(CP)A is a remarkably scalable technique that was demonstrated with attoJoule \cite{Andrekson2008} to Joule \cite{Chekhlov2006} pulse energies and nW to kW average power \cite{Hoppner2015}. Additionally, excellent carrier-envelope phase (CEP) stabilization can be achieved when OPA is carefully implemented. The CEP stabilization via OPAs is often better and more reliable than with broadening/compression sources that rely on active CEP-stabilization of the pump laser system. Exploiting difference-frequency generation (DFG) allows to create pulses with shot-to-shot stable CEP from pump pulses with fluctuating CEP \cite{Baltuska_PRL_2002}. 
Moreover, under particular conditions, the amplification of CEP-stable pulses via OPA does not add significant CEP-noise \cite{Rossi_OptLett_2018}. The coherent combination, also known as \textit{synthesis}, of pulses generated by different sources, each covering a distinct spectral region, offers a way to achieve optical waveforms with multi-octave spanning spectra and durations below a single-cycle. The coherent synthesis was early envisioned for CW-laser sources to redistribute the field intensity within the optical cycle and to create non-sinusoidal waveforms \cite{Haensch_OptComm_1990}. More recently, the synthesis of ultrabroadband pulses from OPAs opened the way to create sculptured waveforms of high intensity and flexible bandwidth \cite{Manzoni_LPR_2015}. The synthesis of $N$ pulses (with similar intensities) covering different spectral regions leads to a pulse whose intensity scales with $N^2$, since not only does the overall pulse energy increase by a factor $N$, but the temporal duration of the synthesized pulse shrinks by the same factor.
The combination of multiple ultrashort pulses of different colors was first achieved by splitting the ultrabroadband output of a hollow-core fiber compressor and compressing different regions individually \cite{Hassan_RevSciInstr_2012}. This approach demonstrated the possibility of shaping complex electric field transients shorter than one optical cycle by controlling the CEP and the delay among the sub-pulses and allowed to achieve for the first time optical attosecond pulses. However, the limited spectral tunability and energy scalability of this approach motivated the development of OPA-based synthesizers \cite{Huang2011,Manzoni2012,Liang2017,Rossi_NatPhot_2020,Kessel2018,Alismail2020,Xue2020}. The high phase stability provided by OPA sources in combination with multi-octave spanning seed generation techniques such as white-light generation \cite{Bellini_OptLett_2000} (WLG) allows to create energy and bandwidth scalable sub-cycle waveforms via the parallel synthesis \cite{Manzoni_LPR_2015} approach in a highly stable manner. The synthesized waveform can be shaped and stabilized in different ways, for instance, by controlling the relative delays (or the relative phases) among the building pulses and their carrier-envelope phases (CEPs).

The increased versatility offered by OPA is achieved at the price of greater complexity than competing schemes. In order to achieve the coherent synthesis of multi-stage OP(CP)As, it is necessary to pay particular attention to the development of the optical and mechanical setup, as well as to implement an elaborate feedback control system.

In this paper, we describe the optomechanical setup and the control system implemented in our recently developed parametric waveform synthesizer \cite{Rossi_NatPhot_2020}. In particular, we focus on different techniques that allowed to achieve exceptional waveform stability and shaping while minimizing the control parameters and simplifying the overall control system. Analytical and numerical modeling of passive CEP-stabilization via DFG and of parametric amplification allowed to determine which conditions maximise phase-stability and to develop an intuitive understanding of the different timing dynamics. This allowed us to simplify the waveform control system with respect to other approaches \cite{Huang2011, Xue2020} and to achieve stable and controllable pulse synthesis.\\
The paper is organized as follows. We start with an overview of the different laser technologies suitable to pump OP(CP)A synthesizers (Sec. \ref{sec_pump_laser}). Afterwards we discuss the dispersion management and beam combination in parallel synthesizers (Sec. \ref{sec_beam_combination}). It follows a discussion of phase stabilization, propagation and control (Sec. \ref{sec_theoretical_description}). In Sec. \ref{sec_technical_implementation} we discuss the most important aspects of the technical implementation. In Sec. \ref{sec_spatial_properties} we describe the spatial properties of the synthesized beam and, in Sec. \ref{sec_streaking} we present results concerning the waveform stability and reproducibility. We conclude with the perspectives of future development and applications of PWS (Sec. \ref{Conclusions}).

\section{Pump Laser Technologies for OPAs}
\label{sec_pump_laser}
In this section, we will describe the main characteristics of laser sources suitable for pumping OP(CP)A-based waveform synthesizers, also called \textit{parametric waveform synthesizers} (PWS). 
Broadband OP(CP)As can be pumped with a wide range of ultrafast laser sources. Over the last decade, a number of high-power ultrafast laser technologies were developed. For instance Yb-doped laser systems based on Yb:YAG thin-disks \cite{Nubbemeyer_OptLett_2017}, Yb-doped fibers \cite{Stark_OptLett_2021}, Yb-YAG slabs/rods \cite{Schmidt_OptExpr_2017,Liu_OSAContu_2020}, Yb:YLF crystals \cite{Demirbas2021} and Yb:KGW (Light Conversion) are viable options to pump parametric amplifiers with kW-level average powers. Nevertheless, for several reasons, Ti:sapphire (Ti:Sa) amplifiers are still broadly used for OPA pumping. Firstly, Ti:Sa amplifiers produce the shortest available pulse durations, down to 20-30 fs for mJ-level commercial systems. A pump pulse duration of $\sim$\,100\,fs is advantageous for OPA since it allows for stable white-light seed generation and simpler OPA output compression. The 800 nm central wavelength is also quite advantageous for pumping common nonlinear crystals such as beta barium borate (BBO) since both the fundamental and its second-harmonic (400 nm) allows for broad phase-matching amplification bandwidth. The output power of commercially available Ti:Sa amplifiers is limited to $\sim$\,15 Watt for room-temperature systems and to $\sim$\,30 Watt for cryo-temperature ones. For many applications, Yb-based systems are usually the best choice when higher power is necessary. One particular advantage of the waveform synthesis scheme based on the parallel combination of OPAs is that it employs multiple pump beams, each pumping a different OPA. This allows to overcome the average power limitations of a single pump laser amplifier by using multiple parallel amplifiers.\\
\begin{figure}
\centering
	\includegraphics[width=\textwidth]{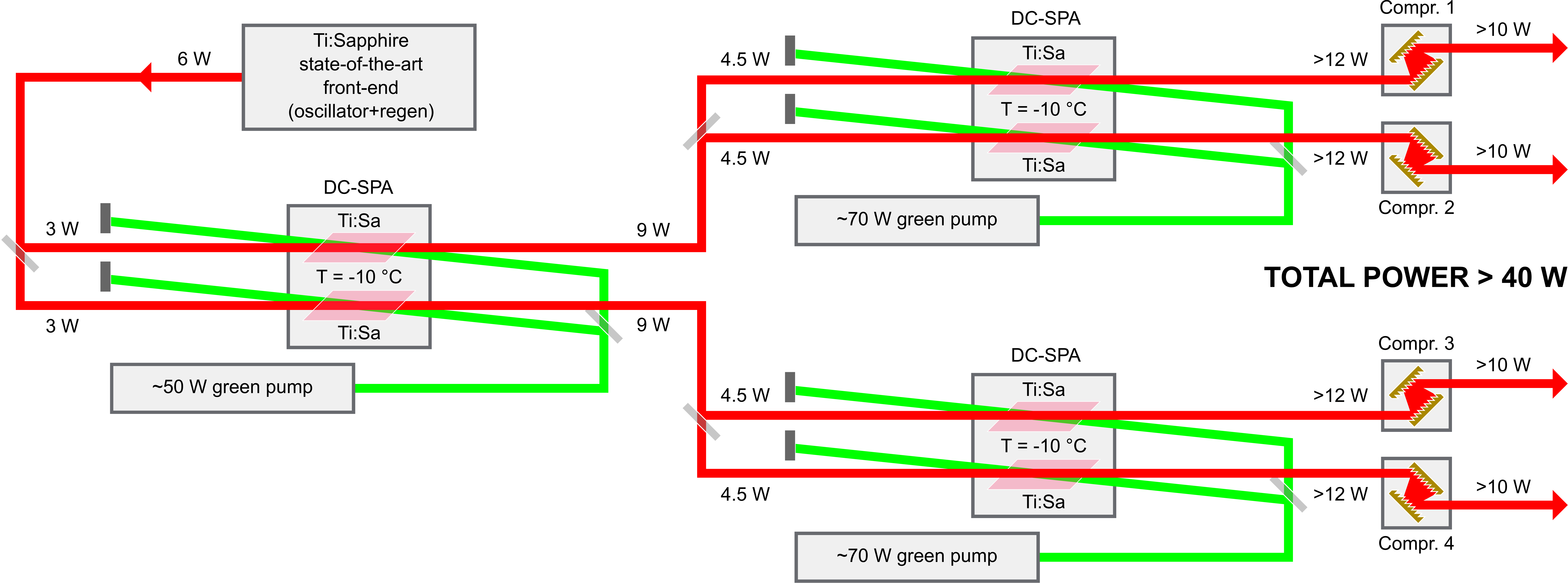} 
	\caption{Example of multi-beam laser system employing separate amplifiers for high-average power pumping of PWS. Scheme of a 4-beam quasi-room temperature Ti:Sa amplifier with $>40$ W output power at 1 kHz in 4 tingly synchronized pulse trains.}
	\label{fig_Ti_Sa_High_Pow}
\end{figure}
Let us consider, for instance, a state-of-art Ti:Sa regenerative amplifier with 6 W average output power (typical for 1 kHz systems). As drafted in Fig. \ref{fig_Ti_Sa_High_Pow}, by splitting and amplifying its output pulses by two consecutive double-crystal quasi-room-temperature (-10$^\circ$ C) single-pass amplifiers, it would be possible to obtain four beams with $>10$ W each after compression ($>40$ W of total power). The power of such a quasi-room-temperature four-beam Ti:Sa system could be even higher, in the order of 70-80 W, by operating at higher repetition rates (3-5 kHz) where common green pump lasers (frequency-doubled Nd:YLF/YAG) reach a higher power. While the multi-beam concept can be implemented with any laser technology, with Ti:Sa it is particularly advantageous since is possible to have single-pass amplifiers with a gain of 2-3 and low B integral. This allows to have just a few meters of non-common beam-path. Jointly with a small compressor size (compressed pulse energy $\leq12$ mJ), this is expected to result in a slight passive timing jitter among output pulses, in the order of few-fs, while preserving an excellent beam quality. Such slight time jitter allows pumping different OPA stages with beams from different amplifiers without the need for active stabilization, as we will show in Sec. \ref{sec_pump-seed_broadband}.\\
This possibility is also intriguing for Yb-doped amplifiers since the longer is the pump pulse duration the higher is the pump-seed jitter admissible in the OPCPAs. For instance, for 300 fs long pump pulses, an rms jitter up to 10 fs would be acceptable. This can be achieved by employing a common oscillator to seed the different amplifiers.\\

Regardless of the specific laser technology, one of the biggest challenges for PWS is the generation of a CEP-stable multi-octave spanning seed pulse. If the pump laser pulses are (sufficiently) CEP-stable, the seed pulse can be obtained directly by spectral broadening of the pump pulse. In case of non CEP-stable pump pulses, intra-pulse difference-frequency generation (DFG) or inter-pulse DFG can be implemented, as a first step, to realize CEP-stable pulses. Inter-pulse DGF, that allows for more freedom in choosing the seed central wavelength and for broader bandwidth, can be realized via white-light generation (WLG) and OPA (see Sec. \ref{sec_theoretical_description}). To produce a shot-to-shot phase-stable WL-filament the pump pulse must exhibit excellent beam quality and rms intensity fluctuations $<1\%$, due to intensity-phase-couplings \cite{Baltuska_JSTQE_2003}. Moreover, in order to have a long-term stable WLG in YAG and Sapphire crystals a pump pulse duration between 100-300\,fs (the shorter the better) is beneficial \cite{Grigutis_OL_2020}. 
Since PWS usually requires several meters of beam propagation, pump beam pointing stability is as well of paramount importance. Beam pointing can be actively stabilized, but since only slow drifts can be effectively compensated, the pump laser should not exhibit high-frequency beam-pointing fluctuations.

\section{Ultrabroadband Beam Combination and Dispersion Management}
\label{sec_beam_combination}
\begin{figure}
\centering
	\includegraphics[width=\textwidth]{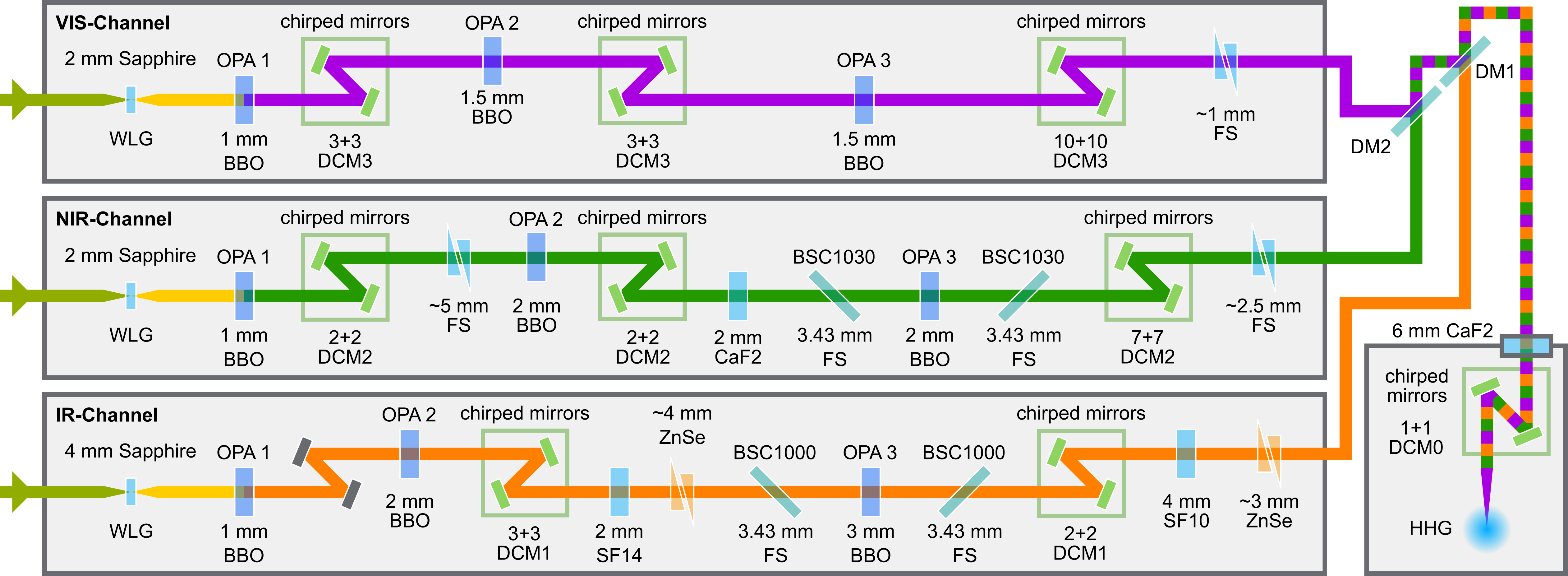} 
	\caption{Dispersion management scheme. The CEP-stable WLG driving beams are shown in light green colour. The dispersion schemes for the IR (orange colour) and NIR (green colour) channels are completed, while the VIS channel (violet colour) it is still under development and its final dispersion scheme might slightly vary. Adapted from \cite{ROSSI2019}.}
	\label{fig_Synth_Optical_Disp}
\end{figure}

\begin{figure}[h!tbp]
\centering
	\includegraphics[width=0.9\textwidth]{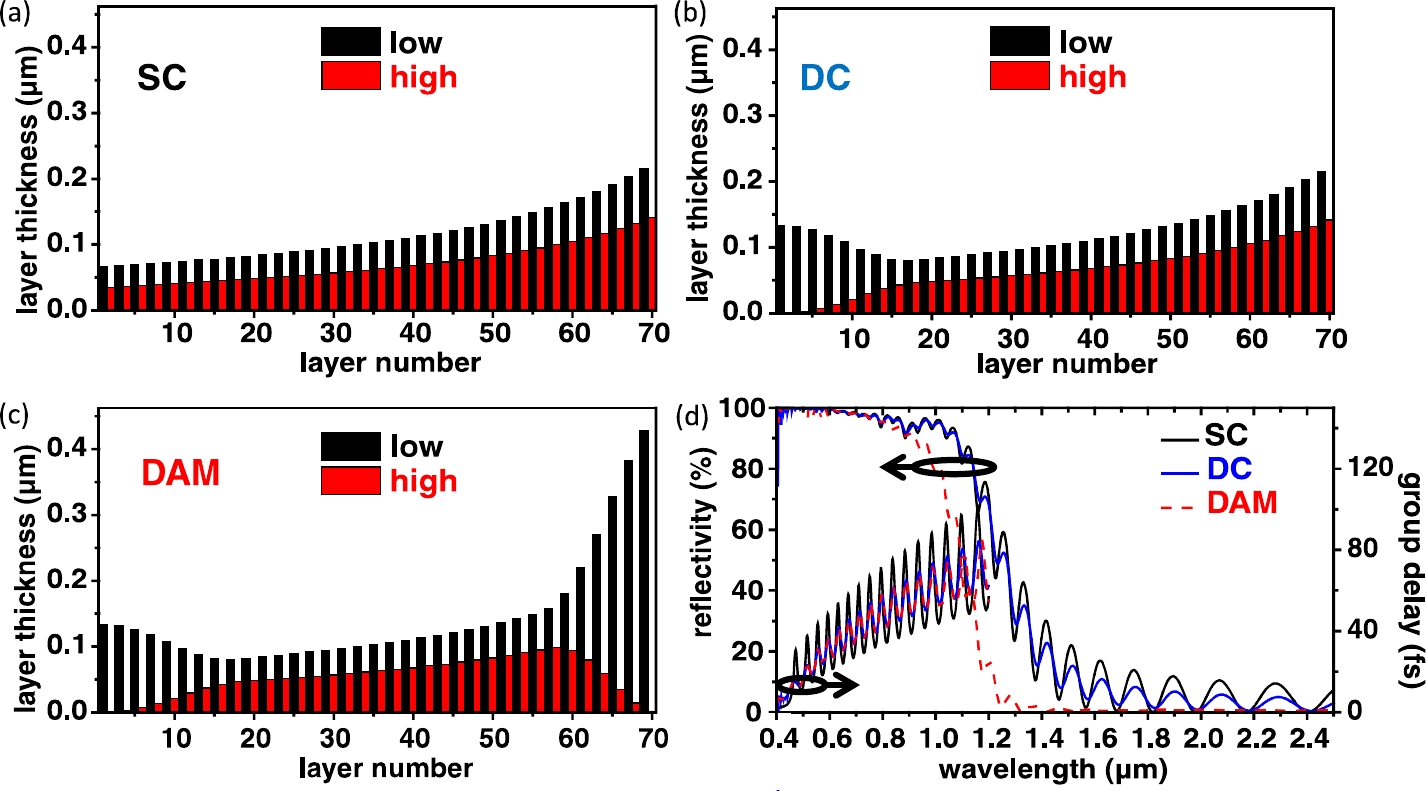} 
	\caption{Schematic of the multi-layer stack of (a) simple-chirped mirrors, (b) double-chirped mirrors, (c) dual-adiabatic-matching mirrors. (d) Reflectivity and group delay of the different mirror structures. Adapted from \cite{Chia_Optica_2014}.}
	\label{fig_DAM}
\end{figure}

\begin{figure}[h!tbp]
\centering
	\includegraphics[width=0.90\textwidth]{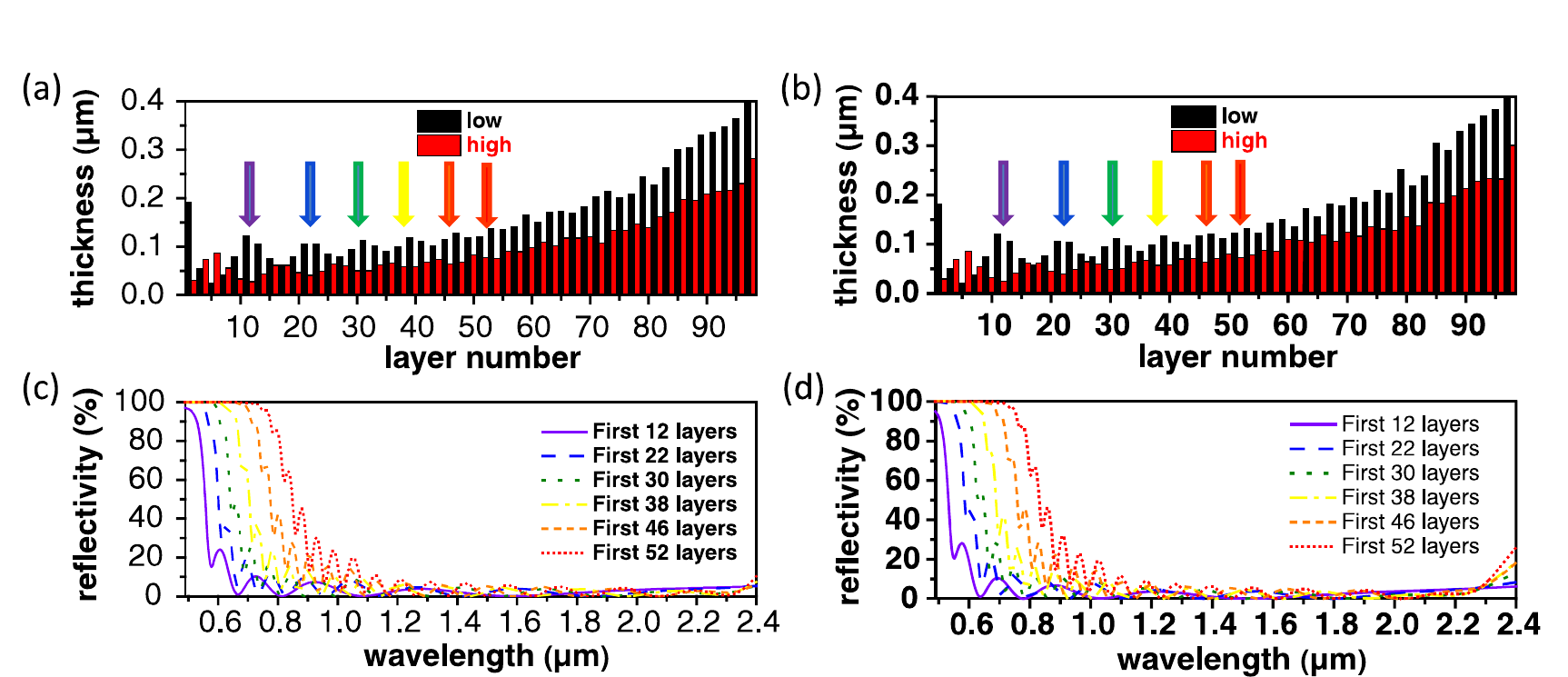} 
	\caption{\textbf{(a)} \& \textbf{(b)} Schematic of the multi-layer stack of the DAM DCM pair. \textbf{(c)} \& \textbf{(d)} Anti-reflection behavior for the longer wavelengths in the cascaded DAM sections indicated by the arrows in the mirror structure in (a) and (b) respectively. Adapted from \cite{Chia_Optica_2014}.}
	\label{fig_UB_DCMs}
\end{figure}

\begin{figure}[h!tbp]
\centering
	\includegraphics[width=0.55\textwidth]{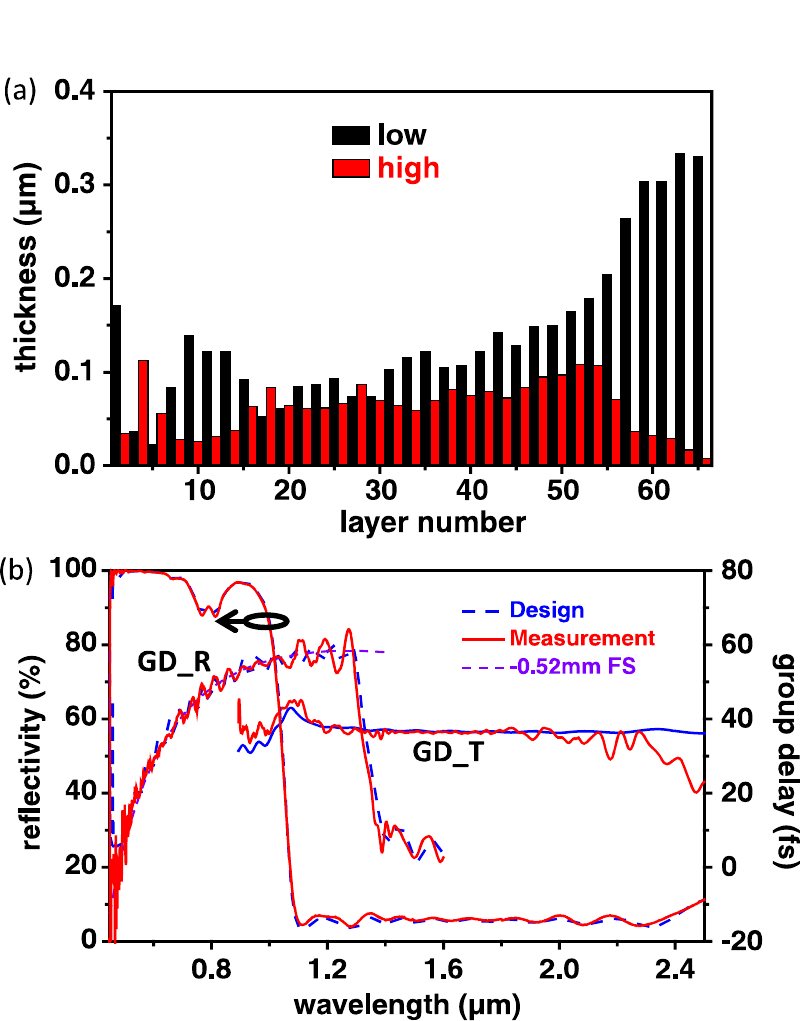} 
	\caption{\textbf{(a)} Structure of a DAM-based dichroic mirror. \textbf{(b)} Designed and measured reflectivity and group delay of the dichroic mirror. The lower transmittance around 0.8 $\upmu$m and the 5\% reflectivity above 1.1 $\upmu$m are intentionally realized to supply weak replicas of the incident pulses to the timing stabilization tools. Adapted from \cite{Chia_Optica_2014}.}
	\label{fig_CDM}
\end{figure}

\begin{figure}[h!tbp]
\centering
	\includegraphics[width=0.60\textwidth]{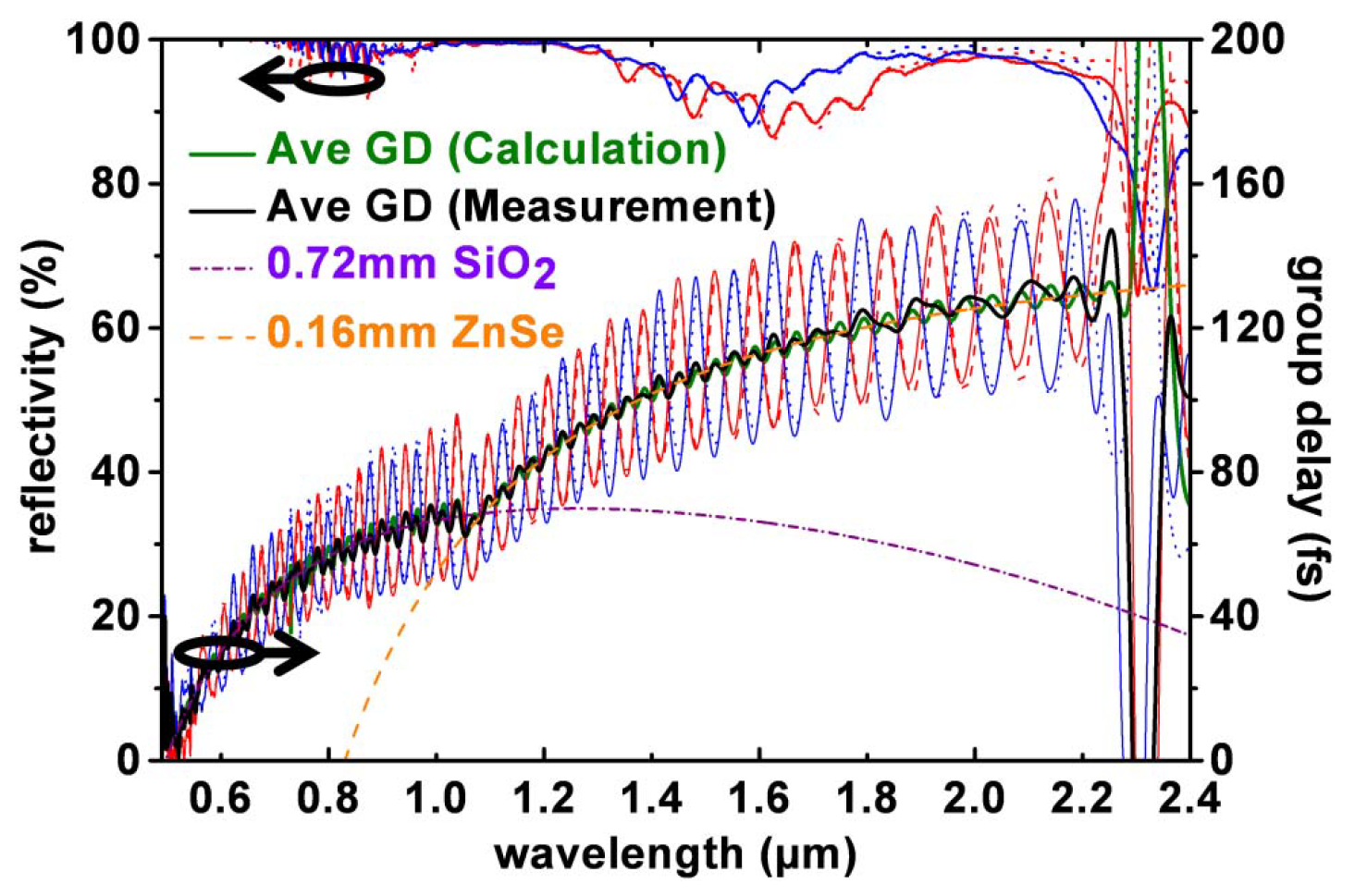} 
	\caption{Reflectivity and group delay of the ultrabroadband DAM DCM pair. The dispersion of the DCM pair compensates a 1.44 mm propagation in fused silica  for the wavelength in the 0.49–1.05 $\upmu$m range, and compensates 0.32 mm propagation in ZnSe for the wavelength in the 1.05–2.3 $\upmu$m range. Adapted from \cite{Chia_Optica_2014}.}
	\label{fig_UB_DCMs_2}
\end{figure}

The generation of ultrabroadband pulses with sub-cycle duration via PWS is enabled by cutting-edge multi-layered optical coatings. In fact, beam combination and dispersion management optics need to support multi-octave spanning spectra with high efficiency. The parallel synthesis scheme, with respect to the serial one, relaxes the constraints on the dispersion management since the dispersion can be managed individually in the different spectral channels, each having < 1 octave of bandwidth. Nevertheless, the dichroic mirrors used to combine the pulses must handle the full synthesized bandwidth. Moreover, it is convenient to use chirped mirrors supporting the full bandwidth to provide the final compression to avoid unwanted nonlinearities during the beam transport into the vacuum beamline towards the experiment. The overall beam combination and dispersion scheme, comprising different dichroic mirrors (DMs), double-chirped mirrors (DCMs), and bulk materials, is shown in fig. \ref{fig_Synth_Optical_Disp}.
For the realization of efficient multi-layered optics capable of dealing with the > 2-octaves bandwidth, a new concept, named \textit{dual-adiabatic-matching} (DAM) structure was developed \cite{Chia_Optica_2014}. The DAM consists in generating an additional double-chirp in the back section of the mirror to provide high transmission for long wavelengths. A comparison of the different chirped-mirror designs is shown in fig. \ref{fig_DAM}.
Thanks to this advance, it was possible to design dichroic mirrors supporting > 2 octaves of bandwidth, capable of high reflectivity and controlled group delay for the short wavelengths and smooth and high transmissivity for the long wavelength, required for efficient splitting and combination of the PWS overall bandwidth. By cascading the DAM structure in the front layers, as an ultrabroadband impedance matching section, it was possible to achieve >2-octave bandwidth double-chirped mirrors (DCMs), shown in fig. \ref{fig_UB_DCMs}.
The average reflectivity of the ultrabroadband DCM pair is $>90\%$, and the calculated peak-to-peak values of the averaged residual GD ripples are controlled to <5 fs over the whole bandwidth. The design of one of the dichroic mirrors and of the final >2-octave spanning DCMs of the PWS are shown in fig. \ref{fig_CDM} and \ref{fig_UB_DCMs_2} respectively.
The final DMs feature a high contrast between the reflected and transmitted beam and an intentionally gradual spectral cross-over region, which allow us to obtain a few percent of energy of the two combined input beams at the secondary port of the beam-combiner. This weak replica of the main beam is used for phase synchronization (see Sec. \ref{Dual In-line Phase Meter}) and monitoring/stabilization of the spectra.
Our current PWS setup will reach its full potential after the implementation of the third spectral channel in the visible range (500-700\,nm), which is already fully supported by our dispersion management scheme and beam combination optics. The overall supported bandwidth spans from 500\,nm up to 2.2 $\upmu$m.
The parallel approach allows to scale this concept to even wider bandwidths. Extending further in the UV is possible although progressively difficult due to a higher susceptibility of layer-deposition inaccuracies, which manifest stronger with shorter wavelengths. In the UV range, material-specific losses of the dielectric coatings need to be avoided, which limits available material pairs. For further extension in the mid-IR these two challenges do not play a major role, and material combinations like Si:SiO$_2$ (>2\,$\upmu$m) or Si:Al$_2$O$_3$ allow to extend the bandwidth up to $\sim$ 7\,$\upmu$m \cite{Razskazovskaya_Optica_2017}. The applied coatings can also exhibit nonlinearities at high pulse intensities, and proper scaling of the mode-size needs to be considered. Light to moderate nonlinearities can be included in the mirror design and pre-compensated. High-bandgap materials such as Nb$_2$O$_5$ might help further to manage nonlinearities \cite{PervakAdOptTec_2014}.\\
Extending the PWS technology towards the mid-IR is also interesting for strong-field experiments such as HHG, where higher photon-energies can be reached \cite{Popmintchev2012}.

\section{Phase Management in a Parametric Waveform Synthesizer}
\label{sec_theoretical_description}
\begin{figure}[!ht]
\centering\includegraphics[width=\textwidth]{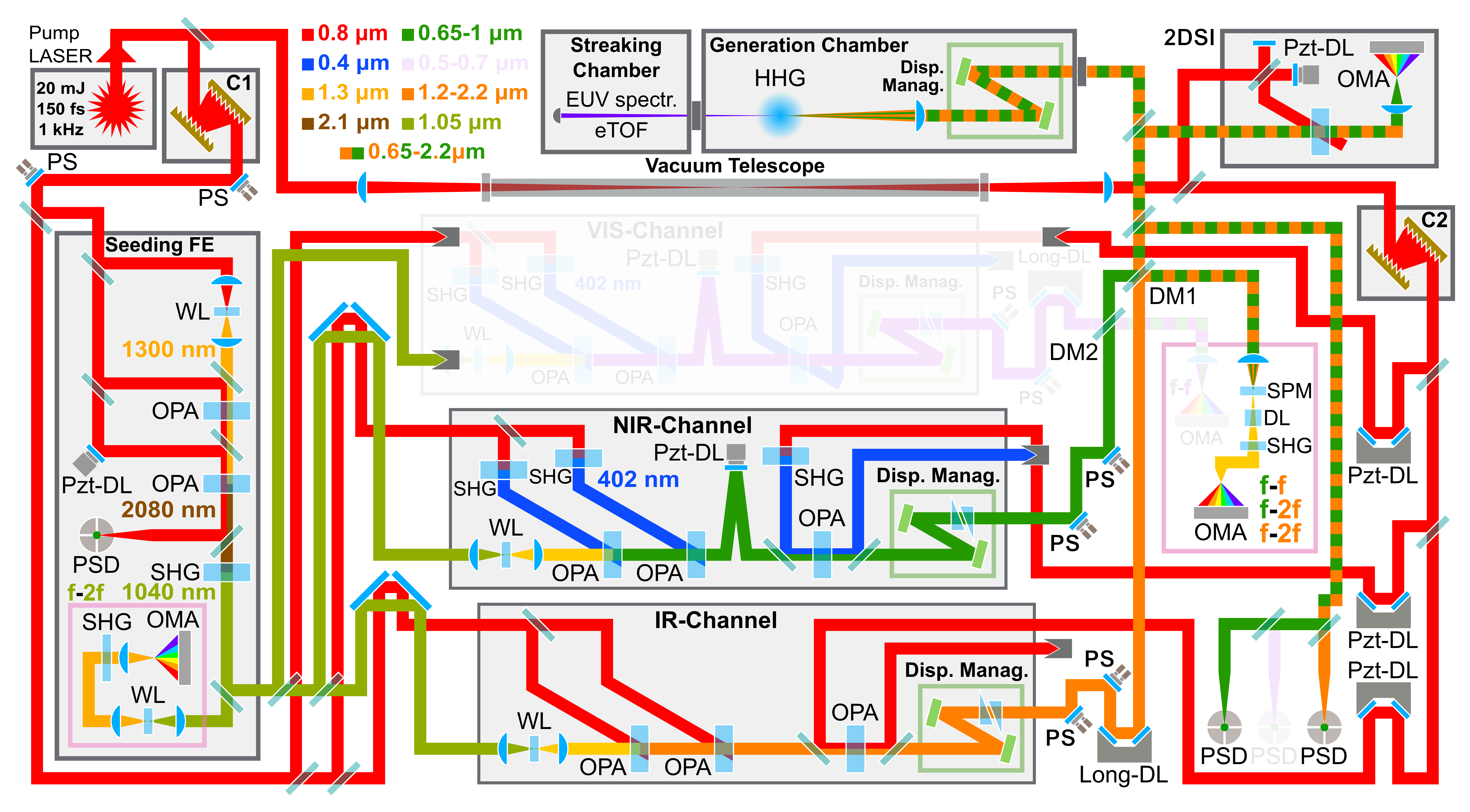}
\caption{Setup of a parallel PWS driven by a Ti:sapphire pump laser at 800\,nm (1\, kHz, 20\,mJ). A fraction of the non-CEP-stabilized laser pulses drives the seeding front-end to derive a passively CEP-stable seed driver  via DFG at 2\,$\upmu$m and its SHG at 1\,$\upmu$m. The SHG pulses drive separate WLG pulses in two parallel OPA-channels in the NIR and IR spectral range with three cascaded amplification stages each and partial recompression after amplification. A third planned VIS spectral channel is shown in the shadowed area. A multi-phase meter at the final beam combination allows detecting all relevant phase-parameters. Long and short-range actuators within the system are used with a distributed feedback to stabilize the synthesis parameters.}
\label{fig_synth_setup}
\end{figure}

In this section, we will analyze the fundamental mechanisms that are at the basis of parallel parametric waveform synthesis (PWS). 
The dynamics that govern each nonlinear process used in a PWS, such as second-harmonic generation, white-light generation, and OP(CP)A have been the subject of numerous studies and are considered well-known. Nevertheless, since the PWS consists of dozens of nonlinear stages, many of which require synchronization between ultrashort pulses, it is helpful to develop a model to describe the impact of timing jitter on the phase of the pulses through the full PWS setup.
This analysis has the task of identifying the essential parameters for controlling the synthesis and clarifying the most critical noise sources by quantifying their impact on the stability of the final synthesized waveform.\\
We will initially focus on the longitudinal properties of the pulses since their stabilization represents the biggest challenge in a parallel PWS. These properties are generically referred as \textit{temporal properties}. The transverse properties (or \textit{spatial properties}) of the synthesis will be discussed in the Sec. \ref{sec_spatial_properties}. Sec. \ref{sec_technical_implementation} will present the most important aspects of the technical implementation of our PWS, in particular, the timing and phase sensors and the low-latency feedback system to actively stabilize and control these longitudinal synthesis parameters.\\
Our PWS scheme (see Fig. \ref{fig_synth_setup}) consists of several cascaded OPA-stages, and the pump-seed temporal synchronization is required in each stage.
In particular, the temporal synchronization among two different ultrashort pulses is required in 9 different components. The number will rise to 13 components once the visible channel is included. These components are the OP(CP)A stages and the dichroic mirrors at the synthesis points (DM1 and DM2 in Fig. \ref{fig_synth_setup}), where the final OP(CP)A outputs are combined.\\
In this section, two questions will be answered: (i) What is the effect of timing jitter among pulses in these different components? (ii) How tight does the synchronization have to be in each component? We know that to achieve a stable waveform, the temporal synchronization required at the synthesis position is in the order of a fraction of the central wavelength, therefore in the hundreds of attoseconds domain. However, this strict timing requirement does not necessarily apply to all components mentioned above. We will demonstrate that while it is necessary to achieve a tight synchronization in some OP(CP)A stages, in other O(PC)PA stages it is not required. This is particularly relevant since it would be extremely complicated to implement a high number of timing sensors and corresponding control actuators and feedback loops. Due to this limitation, it is strictly necessary to analyze the overall timing behavior of the system and find out which is the minimum set of observables and corresponding actuators required to achieve a stable and controllable waveform synthesis.\\
In the following sections, we will start by analyzing the effects of pump-seed jitter in the OPA seeder, then consider the OPA amplifiers, and finally consider the synthesis process.

\subsection{Phase-Stable Seed Generation}
\label{sec_WL-WL}
\begin{figure}[H]
\centering\includegraphics[width=0.85\textwidth]{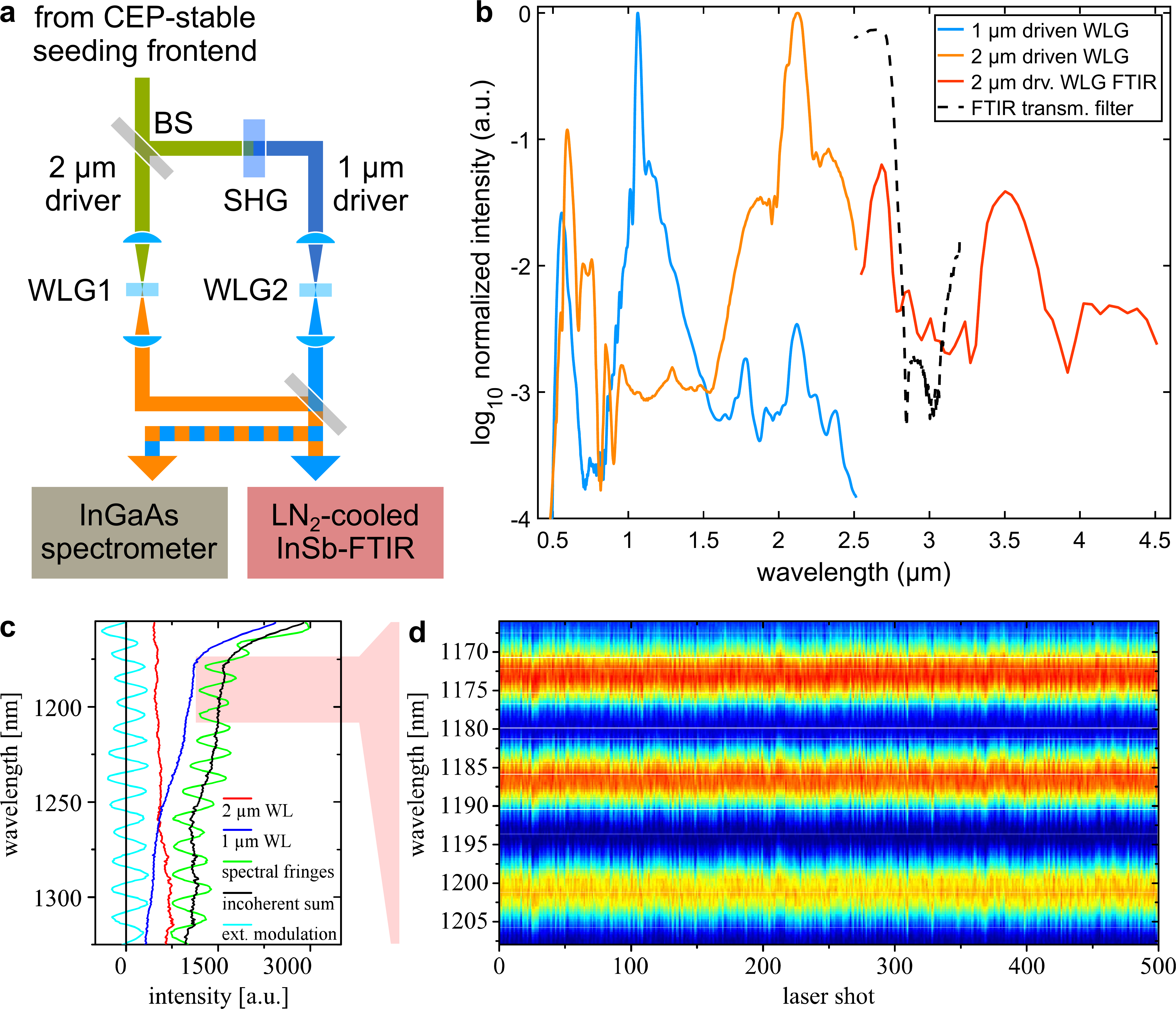}
\caption{\textbf{(a)} Setup for multi-octave wide seed generation via WLG driven by a CEP-stable pulse at 2\,$\upmu$m and its SHG. \textbf{(b)} Optical Spectra from WLG driven at 2\,$\upmu$m (orange/red) and 1\,$\upmu$m (blue) with few $\upmu$J of $\sim$ 120\,fs pulses in bulk media (YAG/sappire). The spectra beyond 2.6\,$\upmu$m are measured with an FTIR using a cryo-cooled InSb-sensor and a lowpass filter (dotted line). \textbf{(c)} Individual WL spectra as recorded with our InGaAs-based spectrometer, as well as the coherent sum (spectral fringes) and the (calculated) incoherent sum. \textbf{(d)} Half-second long section of the interference trace, corresponding to 500 lase shots. Adapted from \cite{Mainz_CLEOUS2016_2016, Mainz_CLEOUS2018_2018}}
\label{fig_WL_bandwidth}
\end{figure}

\begin{figure}[!ht]
\centering\includegraphics[width=0.85\textwidth]{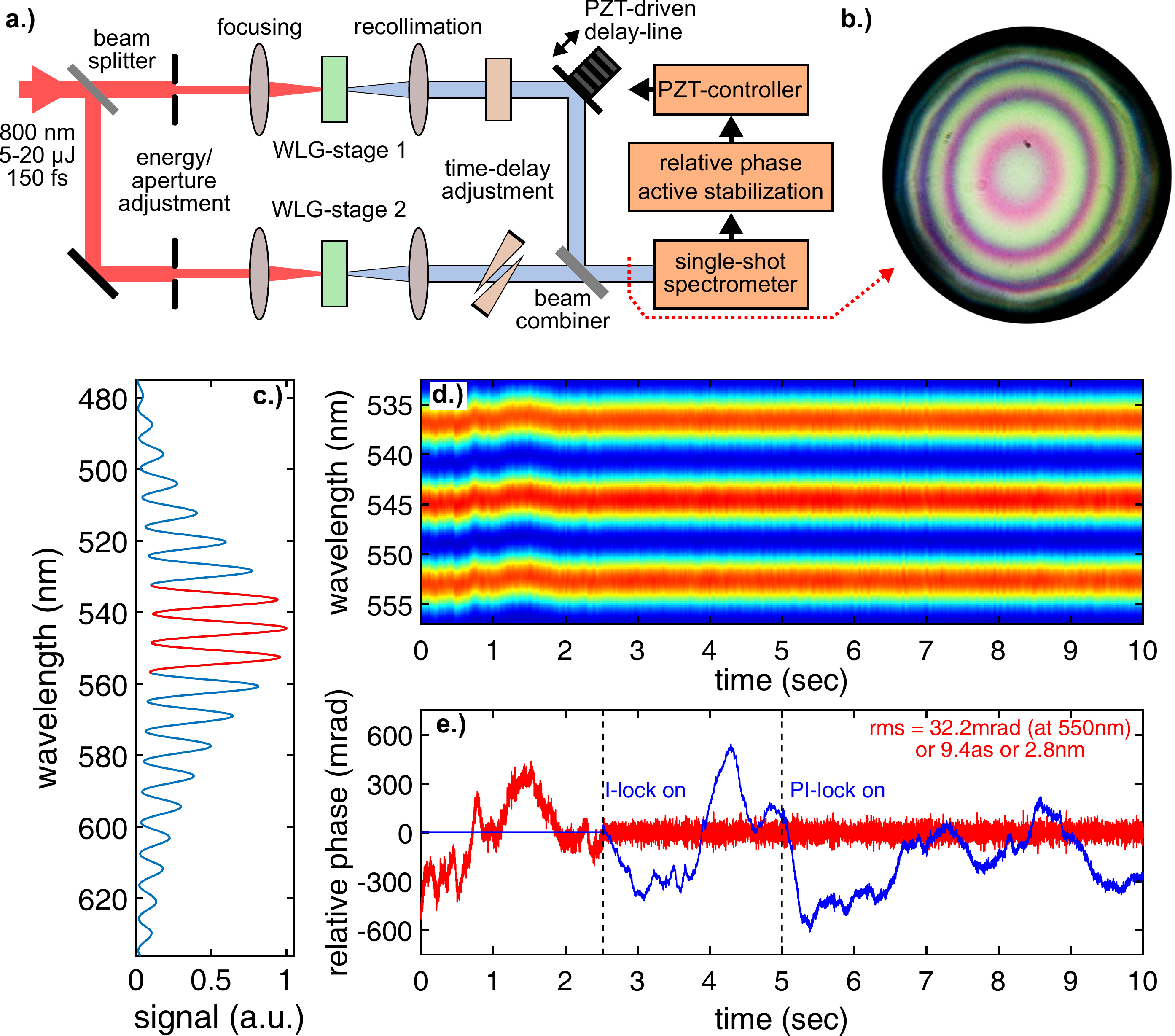}
\caption{Characterization of the phase-stability among separate white-lights. \textbf{(a)} Experimental white-light interferometer for the determination of the relative phase-noise among independent WL. An active stabilization-system eliminates temperature-induced drifts and interferometric noise. \textbf{(b)} Observed spatial fringes between the two WL at the red line marked in (a). \textbf{(c)} Coherent single-hot spectrum of spectra beating between two WLG sources with almost full fringe contrast (95$\%$ modulation). \textbf{(d)} Single-Shot interference trace ( the red marked section in (c) without feedback till 2.5\,s and with activated stabilization after. \textbf{(e)} Retrieved relative phase (red) and delay-line actuation (blue).}
\label{fig_WLWL_fringes}
\end{figure}

\begin{figure}[!ht]
\centering\includegraphics[width=\textwidth]{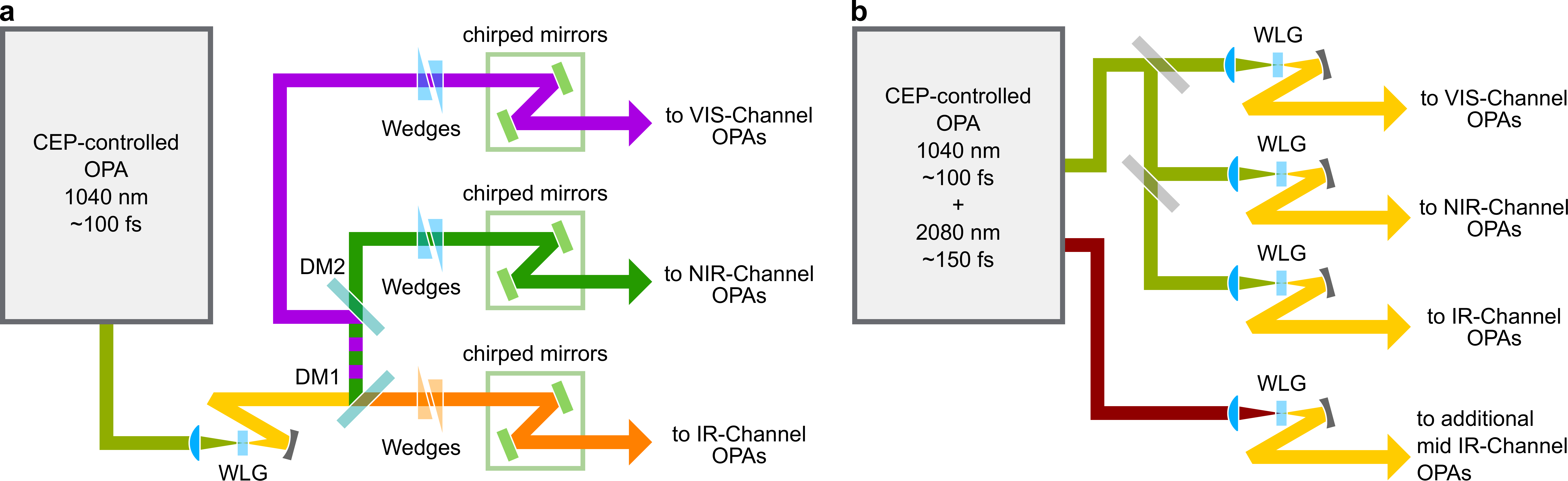}
\caption{Comparison between single-WL \textbf{(a)} and multi-WL \textbf{(b)} seeding schemes. The multi-WL scheme allows to cover a larger bandwidth by driving different WLs with different harmonics, to simplify dispersion management and improve beam quality of the seeding beam due to shorter propagation. Adapted from \cite{ROSSI2019}}
\label{fig_seeding}
\end{figure}

In our PWS, we base the generation of an ultrabroadband and phase-stable seed on the white-light generation (WLG) process\cite{Bellini_OptLett_2000}. With WLG, >2 octaves of seed bandwidth are readily available when driven with sub-ps pulses (see Fig. \ref{fig_WL_bandwidth}). The major drawback of this approach is that the seed pulse has limited energy in the nJ-range, therefore requiring multiple OP(CP)A stages to reach the target energy of 0.1-1\,mJ. On the other hand, WLG allows for multi-octave spanning spectra, tunable from the UV to the Mid-IR range, exhibiting excellent coherence with respect to the driving laser pulse. In particular, the WLG of the PWS, which are driven by a passively CEP-stable pulse at 1040 nm in Sapphire crystals or YAG, can fully span across the 500-2500\,nm region with low intensity fluctuations.
Moreover, WLG exhibits excellent phase stability with respect to its driving pulse \cite{Baltuska_JSTQE_2003}, which allows for maintaining a high CEP-stability when utilizing this spectral broadening technique. The WLG method appears to be advantageous for seeding PWS compared to higher pulse energy seeding techniques like HCFC or MPC since it allows to attain broader bandwidth and better shot-to-shot waveform stability. In our PWS, we implemented separate WLG-stages to prepare an optimized seed for each spectral channel. Earlier observations hint a suitably sufficient phase-stability for such a separate WLG/OPA system \cite{Baum_OptLett_2003}. By adapting the nonlinear material, its thickness, focusing conditions, and pumping intensity, one can optimize the broadened spectra for the different spectral requirements of each spectral channel.\\
To experimentally verify that such a parallel seeding approach is feasible, we characterized the WL phase stability by realizing a Michelson interferometer with two separate WLG stages, one in each arm. Two replicas of  800\,nm pulses drive the WLG stages as shown in Fig. \ref{fig_WLWL_fringes} (a). When both WLG outputs are superimposed, a spectral interference will be observed, and a single-shot every-shot spectrometer quantifies the relative phase-noise among the WLGs. To remove the influences due to slow thermal drifts of the interferometer and air convection, we implemented a low-bandwidth feedback acting on a piezo-actuated delay-line (Pzt-DL) placed in one of the interferometer arms (Fig. \ref{fig_WLWL_fringes} (c)).\\
The residual phase noise is 32-65\,mrad and is mainly correlated to the energy fluctuations of the driving pulse ($\sim$0.5\% rms, \cite{Baltuska_JSTQE_2003}). When both WLG stages use the same crystal material and thickness, the intensity-to-phase coupling coefficients are very similar, and the relative phase noise can be as low as 32\,mrad rms (single-shot, over 10k shots). When different materials or thicknesses are used, the phase noise increases but stays at low absolute levels around 65 mrad rms. The residual rms phase fluctuations are given for different WLG-stages in Tab. \ref{tab:WLG-phase-noise}.\\
\begin{table}[h!]
\begin{center}
\begin{tabular}{l|l|c|c}
			Material 1 & Material 2 & Phase jitter  & Timing jitter  \\
			 &  & (mrad rms (at 550nm)) & as rms \\	 \hline
			2-mm YAG & 2-mm YAG & 32.2 & 9.4 \\ 
			3-mm sapphire & 3-mm sapphire & 41.6 & 12.2 \\
			3-mm sapphire & 2-mm YAG & 51.2 & 15.0 \\ 
			1-mm sapphire & 3-mm sapphire & 64.4 & 18.8 \\
		\end{tabular}
\end{center}
\caption{Overview on relative phase noise (and the corresponding timing jitter) between two separate WLG if driven with 1\,$\upmu m$ and in different materials and thickness. The lowest phase noise is achieved if both WL materials are identical in material and thickness.}	\label{tab:WLG-phase-noise}
\end{table}
To explore the possibility of extending the available seed bandwidth further, we modified the experiment by driving one of the WLs with a 2080\,nm, 100 fs pulse and the second WL with the second-harmonic (at 1040\,nm) of this pulse (Fig. \ref{fig_WL_bandwidth}). In this case, the relative phase stability and the absolute phase (with respect to the seed driving pulse CEP) are quantified. Remarkably the two WLs exhibit good overall phase stability of 132 mrad rms, mainly deteriorating with respect to the previous case due to the additional intensity-to-phase coupling of the second-harmonic generation process involved. In this case, the energy fluctuations of the WLG driver were $\sim$0.5\% for the fundamental pulses and $\sim$1\% for its second harmonic.\\
These results have a few critical consequences for parametric waveform synthesis: (I) there is the possibility to extend the PWS bandwidth by an additional octave in the mid-IR region; (II) a 60 mrad CEP fluctuation of the $\sim$1 $\upmu$m seeder pulse corresponds to an envelope shift with respect to the carrier wave of only 30 as (this will be relevant in Sec. \ref{sec_CEP-RP}); (III) it is possible to operate multiple phase-stable WLs with low phase-noise between each other; therefore it is feasible to seed each synthesizer channel via a dedicated WL.
The last point has a significant impact on the design of the PWS. The possibility of generating separate WL seeds for each spectral channel allows optimizing each of them in terms of spectral intensity/extension, phase, and intensity fluctuations. The seed spectral phase can also be individually optimized by choosing different crystals and thicknesses. Moreover, generating each WL close to the first OPA stage, allows to minimize the WL mode degradation due to extended seed beam transport. The dispersion management scheme is also simplified since the dichroic beam splitters required to split a single seeding WL for the different spectral channels would introduce additional dispersion that needs to be compensated for afterwards. Fig. \ref{fig_seeding} illustrates the single-WL and the multi-WL seeding schemes.

\subsection{Narrowband OPA-Seeder}\label{narrowband}
\begin{figure}[!ht]
\centering
\includegraphics[width=\textwidth]{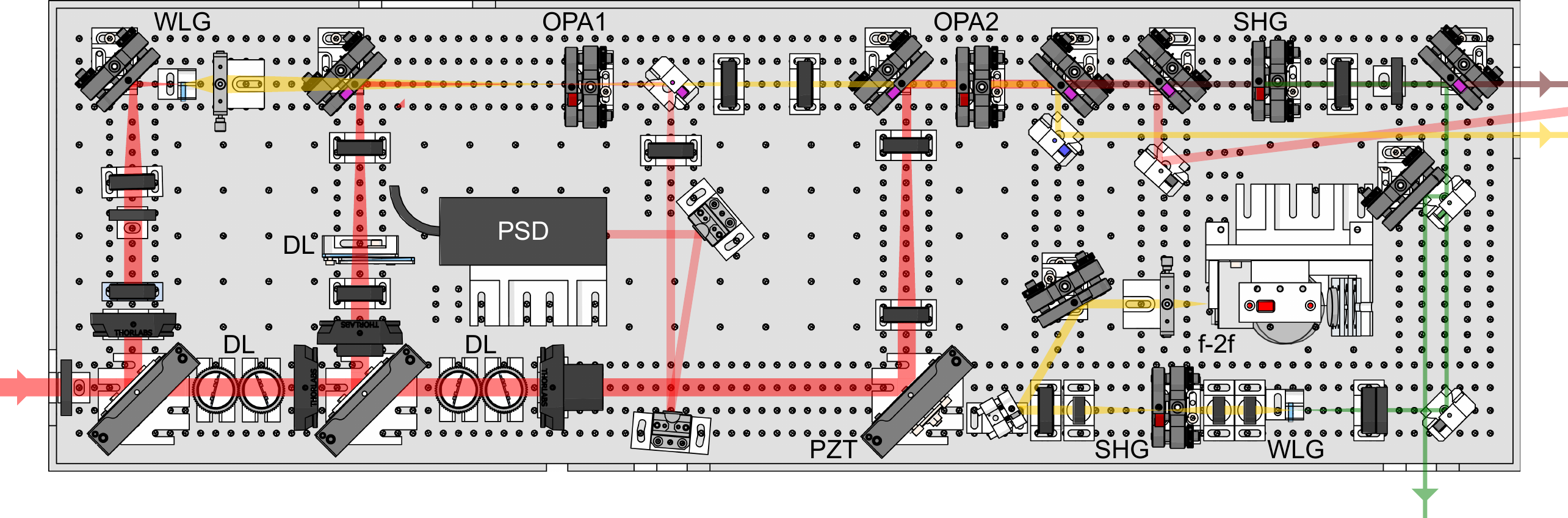} 
\caption{Optical scheme of the CEP-stable OPA seeder. The $\sim$800 nm pump (in red) is split in three replicas, the first one driving the WLG (in yellow), the others pumping the first and second OPA stages (OPA1 and OPA2). The fist stage pump after amplification is directed to the pointing stabilizer detector (PSD), the corresponding actuators are placed outside the breadboard (not shown). Wedges act as optical delay lines (DL) in transmission and are used to fine tune the pump-seed delay in each OPA stage. The pump-seed delay in the second amplification stage (OPA2) can be actuated by the piezo-driven delay line (PZT), allowing for further active stabilization and control of the CEP of the idler beam. The CEP-stable idler (brown colour) it is frequency-doubled to obtain the 1040 nm CEP-stable pulse that drives the seed generation in the PWS spectral-channels. Its CEP stability is measured locally by means of an f-2f interferometer.}
\label{fig_CEP_top_beam}
\end{figure}

As discussed in the previous section, the generation of a broadband and phase-locked seed relies on the WLG process. However, to generate a seed with stable CEP, the WL itself needs to be driven by a CEP stable pulse. In most cases, high-power laser amplifiers can not be CEP-stabilized, or if so, they do not reach the stability necessary for pulse synthesis. For this reason, the generation of CEP-stable pulses via DFG appears to be the most reliable approach.\\
The simplest option is to make intra-band DFG from the pump laser oscillator pulse if it is too narrowband by broadening it beforehand. However, this method can only produce a CEP-stable seed in the IR and mid-IR. Additional nonlinear processes (ex. SHG) would be needed to extend the seed in the visible and near-IR region. In order to control the CEP and further stabilize it, it is necessary to introduce dispersion before or after the DFG stage. In order not to affect the relative arrival time of the seed with respect to the pump in the following OPAs, an isochronic CEP compensator needs to be implemented \cite{Gorbe2008}.\\
Alternatively, the DFG can be done in a WL-seeded OPA. This approach allows obtaining a tunable CEP-stable seed pulse that can be further broadened via WLG.
To obtain both shot-to-shot and also long-term stability, WL generation also requires driving pulses with excellent energy stability, a Gaussian beam profile, and preferably a pulse duration in the 50-300 fs range.\\
In this section, we will present the main technical aspects of the implementation of our OPA-based seeder system.\\
Our seeder (Fig. \ref{fig_CEP_top_beam}) consists of a WL-seeded two-stage OPA, pumped with 800\,nm and generating a signal at 1300\,nm (idler at 2080\,nm). Both stages, based on BBO crystals (cut at 25.9$^\circ$), exploit the \textit{Type-II oee} phase-matching geometry to achieve the condition in which, inside the crystals, the signal and the idler pulses tend to walk away from the pump pulse in opposite directions, that is $\delta_{s-p}\cdot\delta_{i-p} < 0$ where $\delta_{j-k} = 1/v_{g_j} - 1/v_{g_k}$. The seed pulse, faster than the pump pulse, enters the crystal after the pump so that the two pulses overlap temporally after a short propagation. Once overlapped, the parametric amplification begins, and the generated signal and idler pulses remain locked, due to the gain, to the pump pulse for a distance much longer than both pump-signal and pump-idler pulse splitting lengths ($l_{s/i-p} = |\tau/\delta_{s/i-p}|$). Once the OPA-gain drops due to pump depletion, the three pulses separate from each other because of the different group velocities and they cease interacting. By choosing the proper crystal length (in our case, 2.5 mm in the first stage, 4 mm in the second stage), it is possible to reach a condition for which the output signal and idler pulse energy depend neither on the input seed energy nor on the pump-seed arrival time difference fluctuations, but exclusively depend on the input pump pulse energy. In this OPA configuration, discussed for the first time in \cite{Manzoni_JOpt_2016}, the signal and seed pulse energy fluctuation can be as low as the pump one.
As discussed in the next section, the idler pulses generated with this OPA are passively CEP-stable. After the second stage, the idler pulses are separated from the pump and signal and frequency-doubled (in a \textit{Type-I} BBO with 0.5 mm thickness to preserve beam quality at the expense of conversion-efficiency) to obtain 1040 nm pulses, whose wavelength is particularly suited for driving the WL seeds for the PWS. The broadband OPA amplifiers of the spectral-channels cover the 500-700 nm range (VIS-channel), the 650-1000 nm range (NIR-channel), and the 1200-2200 nm range (IR-channel), leaving a gap in the 1000-1200 nm region. Since the WL driver central wavelength is contained in this gap, the distorted regions of the WL spectrum around the pump wavelength are not utilized but only its spectral wings, that exhibit a smooth spectral intensity/phase.\\
After discussing the OPA scheme that minimizes energy fluctuations we will briefly describe its technical implementation that plays a fundamental role in achieving stable long-term operation, excellent beam profile and high CEP stability. To achieve long-term stability, we minimized the effects of the environment (temperature changes and air fluctuations) on the setup.
Two measures drastically increase the stability and consist of reducing the beam path length and minimizing the number of reflecting optics in favor of transmissive elements.
The compact Mach–Zehnder-type geometry (see Fig. \ref{fig_CEP_top_beam}) achieves those goals and further helps to keep the multiple pump beams roughly synchronized to the signal path. To minimize the effect of dispersion through the transmissive optics, we used SiO\textsubscript{2} optics, whose zero-dispersion wavelength ($\sim$1.3 $\upmu$m) coincides with the signal central wavelength.\\
To further reduce the environmental influences on the optical system, we implemented a temperature stabilized ($\sim$3 mK rms) breadboard, described in Sec. \ref{sec_Optomechanical_design}, and fully enclosed the setup including windows for the input and output beams. Moreover, a pointing stabilizer (Aligna, TEM Messtechnik) was added to fix the input pump beam direction with respect to the optical breadboard itself (PSD in Fig. \ref{fig_CEP_top_beam}).\\
The quality of the beam profile of the idler second-harmonic also plays a vital role since the stability of the WL dramatically depends on it. Furthermore, the SHG pulses at 1040 nm have to propagate for a few meters before reaching the WLG stage of some of the spectral channels (as discussed at the end of the last section); therefore, to avoid diffraction, the beam profile needs to be Gaussian. To avoid angular dispersion and achieve an excellent idler beam profile, we adopted a perfectly collinear OPA geometry and used custom dichroic mirrors (produced by Laseroptik) to combine and split the different beams. 

\subsection{Pump-seed Timing Effects on CEP for Narrowband OPAs}\label{sec_pump-seed_narrowband}
We will now discuss how the idler CEP stabilization and control works in our OPA seeder. The so-called \textit{passive} CEP stabilization process occurring in OPAs, first observed by \textit{Baltuška et al.} in 2002 \cite{Baltuska_PRL_2002}, is explained by the fact that the idler pulse is generated during the OPA process by the difference-frequency generation (DFG) between the pump pulse and the seed (signal) pulse. Provided that a replica of the pump pulse generates the seed pulse through a process that preserves the phase coherence (as for WLG), the shot-to-shot CEP variations of the pump will be transferred identically to the seed pulse. In this case, during the DFG process, if we assume that the pump-seed temporal overlap does not vary from shot to shot, the CEP fluctuations of the seed pulses are subtracted from the CEP fluctuations of the pump pulse, resulting in an idler pulse whose CEP is constant for every shot. However, the assumption that the pump-seed temporal overlap in the OPA does not fluctuate is not valid in general. The effects of the pump-seed arrival time difference (ATD) fluctuations on the CEP of the idler pulse were studied in \cite{Rossi_OptLett_2018}. In this section, we will summarize the main results reported in that study concerning narrowband OPAs such as the seeder of our PWS.\\
The influence of a shot-to-shot pump-seed ATD variation named $\Delta T$ can be found analytically for the two extreme cases $\tau_{seed} \gg \tau_{p}$ and $\tau_{p} \gg \tau_{seed}$. When $\tau_{seed} \gg \tau_{p}$, the CEP variation of the idler is:
\begin{equation}
\Delta\Psi_{i} = -\omega_{seed}\Delta T,
\label{idler_CEP_narrowband_1}
\end{equation}
otherwise, when $\tau_{p} \gg \tau_{seed}$, the CEP variation of the idler is:
\begin{equation}
\Delta\Psi_{i} = -\omega_{p}\Delta T.
\label{idler_CEP_narrowband_2}
\end{equation}
Here $\omega_{seed}$ and $\omega_{p}$ are the central angular frequencies of the seed and pump pulses. Since in an OPA it is always true that $\omega_{p} > \omega_{seed}$, a temporal fluctuation $\Delta T$ will affect less the stability of the idler CEP if $\tau_{seed} \gg \tau_{p}$. This result also suggest that the idler CEP can be controlled by acting on the pump-seed ATD. In our OPA seeder (where $\nu_{seed} \approx 230$ THz), the CEP of the idler shifts by $\pi$ for a $\Delta T \approx 2.2$ fs.\\
In light of the above, a 12 nm band-pass filter centered at 1300 nm is placed between the first and the second stage of our OPA seeder. Such filter narrows the bandwidth of the amplified signal emerging from the first OPA-stage such that $\tau_{seed} > \tau_{p}$ and also served as a block for the first stage idler, preventing double seeding of the second stage. This helps to minimize the CEP noise induced by fluctuations of the beampaths in the OPA.
A Pzt-DL (PZT in Fig. \ref{fig_CEP_top_beam}) consisting of a mirror mounted on a piezo-actuator is added to adjust the pump-seed delay in the second OPA stage, therefore enabling control over the idler CEP. The delay line is introduced on the beampath of the pump for convenience reasons. Our OPA seeder is also equipped with an f-2f interferometer and a spectrometer enabling single-shot CEP detection of the final output pulses at 1040 nm at the full repetition rate (see Sec. \ref{sec_control_system}).  

\subsection{Broadband OPAs}
\label{broadband}
In sections \ref{sec_WL-WL},\ref{narrowband} and \ref{sec_pump-seed_narrowband} we described the key ingredients for the generation of multi-octave spanning WL seeds with stable and controllable CEP. In this section, we will discuss how to amplify the nJ-level seeds up to the mJ-level while preserving phase coherence. As mentioned in Sec. \ref{sec_beam_combination}, the WL seed spectrum, spanning 2.5 octaves across the visible and IR range, can be almost entirely amplified by means of three different ultrabroadband OPA configurations. The ultrabroadband amplification bandwidth of all these OPAs is obtained via \textit{group-velocity matching} of signal and idler pulses. In fact, when $v_{gs} = v_{gi}$, the first order term of the amplification bandwidth \cite{Cerullo_RevSciInstr_2003}:
\begin{equation}\label{eq:bandaguad}
\Delta\omega_G = \alpha | v_{g_i}^{-1} - v_{g_s}^{-1}|^{-1} + \beta| GVD_s + GVD_i|^{-\frac{1}{2}} + ...
\end{equation}
tends to infinity ($\alpha$ and $\beta$ contain OPA parameters). The group-velocity matching condition can be achieved both via \textit{non-collinear OPA} configuration, as in the VIS-channel, or via \textit{degenerate OPA} configuration, as in the NIR and IR channels.

\subsection{Pump-seed Timing Effects on CEP, Broadband Stretched/Compressed OPAs}\label{sec_pump-seed_broadband}
In Sec. \ref{sec_pump-seed_narrowband}, we discussed the effect of pump-seed delay on signal and idler CEPs in a narrowband OP(CP)A, where the seed pulse is transform-limited (TL). In broadband OP(CP)As, as those implemented in the PWS spectral channels, the CEP dependency on pump-seed delay is quite different. In this case, before entering the nonlinear medium, the broadband seed pulse is usually stretched in order to fit the high-gain temporal window induced by the pump pulse (generally $\tau_{p} > 10\cdot\tau_{seed_{TL}}$) and extract more energy from it. After amplification, the signal (or idler) pulses are compressed close to TL.\\ 
In order to describe the CEP dependency on the pump-seed delay analytically, we can extend the simple model developed in \cite{Rossi_OptLett_2018} to the chirped seed case. To do this we start with the seed pulse:
\begin{equation}
E_{seed_0}(t) = e^{-\left(\frac{t}{\tau_{seed_0}}\right)^2} e^{i(\omega_{seed}t + \phi_{seed})}.
\end{equation}
The stretcher can be modeled in the frequency domain by multiplying the seed pulse by:
\begin{equation}
S(\omega) = e^{iGD\omega}e^{-i\left[CD\omega_{seed} + GD(\omega-\omega_{seed}) + \frac{1}{2}GDD(\omega-\omega_{seed})^2\right]},
\label{stretcher_w}
\end{equation}
where CD, GD and GDD are the \nth{0}, \nth{1} and \nth{2} order dispersion of the stretcher. The seed pulse in time domain after the stretcher is then:
\begin{equation}
E_{seed_1}(t) = A_1 e^{-\left(\frac{t}{\tau_{seed_1}}\right)^2} e^{i\left[\phi_{seed} + (GD-CD)\omega_{seed} - \frac{1}{2}\arctan\left(\frac{2GDD}{\tau_{seed_0}^2}\right)\right]} e^{i\left(\omega_{seed}t + \frac{2GDD}{\tau_{seed_1}^2\tau_{seed_0}^2}t^2\right)},
\end{equation}
where $\tau_{seed_1} = \tau_{seed_0}[1 + 4(GDD/\tau_{seed_0}^2)^2]^{1/2}$ and $A_N$ ($N = 1,2,...$) are a suitable amplitudes.\\
We now assume that the signal pulse generated during OPA is a (higher intensity) copy of the stretched seed pulse multiplied by a shorter pump pulse ($\tau_{seed_1} > \tau_{p}$). In the opposite limit case ($\tau_{p} \gg \tau_{seed_1}$) it is trivial to show that the pump-seed delay has no effect on signal CEP. If $\tau_{seed_1} > \tau_{p}$, the signal inherits the pump envelope, whose temporal peak (the arrival time) shifts according to the pump-seed delay T:
\begin{equation}
E_{s_0}(t) = A_2 e^{-((t - T)/\tau_{p})^2} e^{i(\phi_{seed} + (GD-CD)\omega_{seed} - \frac{1}{2}\arctan(\frac{2GDD}{\tau_{seed_0}^2}))} e^{i(\omega_{seed}t + \frac{2GDD}{\tau_{seed_1}^2\tau_{seed_0}^2}t^2)}.
\end{equation}
The CEP of the signal pulse is equal to the phase at the peak of its envelope, therefore at $t = T$:
\begin{equation}
\Psi_{s_0} = \phi_{seed} + (GD-CD)\omega_{seed} - \frac{1}{2}\arctan(2GDD/\tau_{seed_0}^2) + \omega_{seed}T + \frac{2GDD}{\tau_{seed_1}^2\tau_{seed_0}^2}T^2.
\end{equation}
With respect to the narrowband TL case, where $\Delta\Psi_{s}(T) = \omega_{seed}\Delta T$, here the signal CEP gains an additional quadratic dependence on the pump-seed delay due to the second order dispersion ($GDD$).\\
Let's now see what happens once the signal pulses are compressed. To this end we multiply the signal pulse (in the frequency domain) by $S(\omega)^{-1}$, that is a compressor ideally matched to the stretcher $S(\omega)$. Once back in the time domain, the signal pulse is:
\begin{equation}
E_{s_1}(t) = A_3 e^{-\frac{[4GDD^2T + (\tau_{seed_0}\tau_{seed_1})^2(t - T)]^2}{\tau_{seed_1}^2(16GDD^4 + \tau_{seed_0}^4\tau_{seed_1}^4 + 4GDD^2(\tau_{seed_0}^4 - 2\tau_{seed_0}^2\tau_{seed_1}^2)}} e^{iB},
\label{signal_compressed}
\end{equation}
where $B$ contains all the phase terms. We can find out the temporal position of the peak of the Gaussian envelope by looking where the argument of the exponential equals zero, that is for:
\begin{equation}
t_{peak} = \left(1 - \left(\frac{2GDD}{\tau_{seed_0}\tau_{seed_1}}\right)^2\right)T .
\end{equation}
By substituting $t_{peak}$ into the full expression of $B$ in Eq. \ref{signal_compressed} we can finally obtain the CEP of the signal pulse after compression:
\begin{equation}\label{signal_CEP}
\Psi_{s_1} = \phi_{seed} + \left(\frac{\omega_{seed}}{R^2}\right) T + \left(\frac{2GDD}{\tau_{seed_0}^4 R^4}\right)T^2,
\end{equation}
where $R = \tau_{seed_1}/\tau_{seed_0}$ is the stretching (and compression) ratio. This expression allows us to observe that for large stretching ratios $\Delta\Psi_{s}(T) \rightarrow \phi_{seed}$. The condition $R \gg 1$ is verified in any broadband OP(CP)A, therefore this expression for signal CEP has general validity.\\
We conclude that the CEP of the signal pulse is not influenced by the pump-seed timing fluctuations, even when the stretched seed pulses are longer or similarly long with respect to the pump pulses, provided that the signal is later re-compressed close to TL. In case of TL seed/signal pulses, that is for $R \rightarrow 1$, we recover the same expression found in \cite{Rossi_OptLett_2018} for the narrowband TL case with $\tau_{seed} \gg \tau_{p}$.\\
The same derivation can be applied to the idler pulse, with the difference that the idler carrier is given by the difference between pump and seed carriers. For the idler pulse we obtain that:
\begin{equation}
\Psi_{i_1} = \phi_{p} - \phi_{seed} + \left(\frac{\omega_{i}}{R^2} - \omega_{p}\right)T - \left(\frac{2GDD}{\tau_{seed_0}^{4}R^{4}}\right)T^2\\
= \phi_{p} - \omega_{p}\left(1-\frac{1}{R^{2}}\right)T - \Psi_{s_1},
\label{idler_CEP_broadband}
\end{equation}
where $\omega_i = \omega_p - \omega_{seed}$. Differently from the signal case, the pump-seed delay contribution to idler CEP after re-compression does not vanish for large stretching ratios ($R \gg 1$):
\begin{equation}
\Psi_{i_1} = \phi_{p} - \phi_{seed} - \omega_{p}T,
\end{equation}
in agreement with idler CEP expression (\ref{idler_CEP_narrowband_2}) valid for $\tau_{p} \gg \tau_{seed}$. Consistently, when we consider narrow-band seed pulses with no dispersion ($GDD = 0$ and $R = 1$), the Eq. \ref{idler_CEP_broadband} leads to Eq. \ref{idler_CEP_narrowband_1}, that was found for $\tau_{seed} \gg \tau_{p}$.
\begin{figure}[H]
\centering
	\includegraphics[width=1\textwidth]{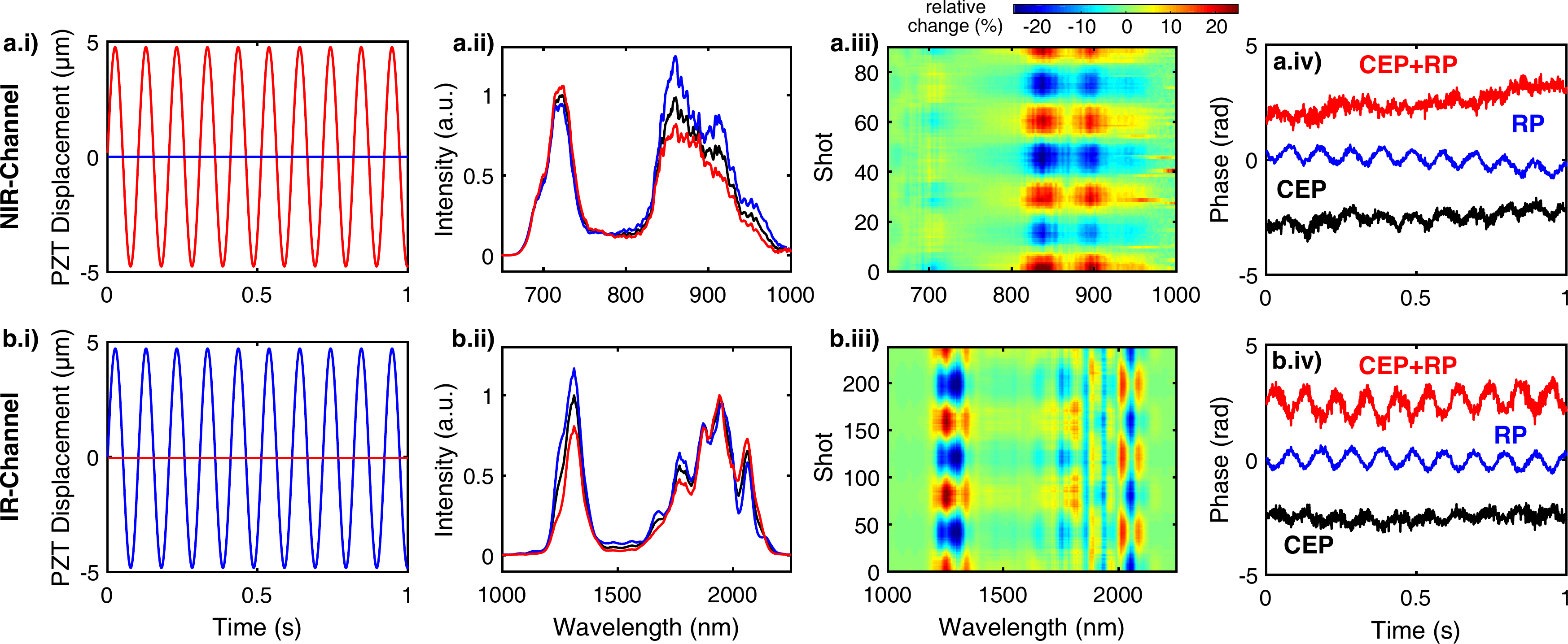} 
	\caption{Pump-seed ATD effects on the synthesis channels for \textbf{(a)} the NIR-channel and \textbf{(b)} the IR-channel. \textbf{(i)} Shows the sinusoidal ATD modulation with 60\,fs$_{pp}$. \textbf{(ii)} Average spectra (black) and extrema of spectral modulation (red/blue). \textbf{(iii)} Trace of the spectral shift at the synthesizer channel output during pump-seed modulation. \textbf{(iv)} Effects on the measured phase of the CEP+RP (red), the RP (blue) and the CEP-effect (black) attained by subtraction.}
	\label{fig_pump_seed_spectra}
\end{figure}
As a practical example, we consider the influence of pump-seed temporal drifts on the signal CEP of the NIR channel of the PWS. In Equation \ref{signal_CEP} we substitute $R = 10$, $\omega_{seed} = 2.36 \times 10^{15}$ rad/s and $T = 1$ fs to the second term, while we can neglect the third them. This leads to a signal CEP shift (after compression) of 23.6 mrad for a 1 fs change in pump-seed arrival time difference. To experimentally verify this calculation, we induced a large change in the pump-seed delay of the laser OPA amplifiers of both NIR and IR channels. The changes induced in the amplified spectra of both NIR and IR pulses are shown in Fig. \ref{fig_pump_seed_spectra} (a.ii, b.ii) and (a.iii, b.iii), while the corresponding phases are shown in Fig. \ref{fig_pump_seed_spectra} (a.iv, b.iv). The phase change is in the range of 1-1.5 rad for a $\sim$60 fs delay, corresponding to 16.7-25 mrad for 1 fs, remarkably close to the value obtained from the previous model. In the PWS last stage amplifiers, we observed pump-seed shot-to-shot fluctuations $<2$ fs rms, corresponding to small CEP changes. Additional larger drifts, up to $\sim$10 fs, can occur slowly, over several tens of minutes, due to temperature changes in non-stabilized parts of the setup, such as when opening (and leaving open) the PWS box. In the next session we will quantify the impact on the synthesized waveform of non-common mode CEP drifts among the two (and three) spectral channel pulses (see Fig. \ref{synth_2ch_1+2} \& \ref{synth_3ch_1+2}).\\
From this analysis, we conclude that, in the broadband OPA amplifiers of the spectral channels of the PWS, it is not necessary to actively stabilize the pump-seed temporal jitter to maintain CEP (and RP) stability. This statement is valid as long as: (i) we are interested only in the signal pulses from the spectral channels, and (ii) the pulses are utilized in the experiment once compressed. Both these conditions are met in our PWS. This has interesting implications for next-generation OPCPA synthesizers. The seeding laser could be synchronized electronically to a high-energy/power amplifier to pump the different OPCPA stages without introducing significant phase noise. For instance, for a $\geq$1.5 ps pump duration and an $R = 100$ (assuming a 10 fs TL duration of the seed), a pump-seed jitter up to 100 fs (peak-to-peak) would not significantly impact the CEP stability.
In our PWS setup we implemented a slow-feedback that stabilize the slow drifts (0.1-1 mHz bandwidth) of the pump-seed ATD in the last OPA stage, that can reach few tens of fs. This is mainly due to the thermalization of the external compressor (C2 in Fig. \ref{fig_synth_setup}). These slow drifts have no impact on the CEP stability but leads to small shifts in the OPA output spectra. The simple stabilization scheme is based on the calculation of the center of mass of the measured OPA spectra (acquired every $\sim$\,1 minute) and is stabilized by moving the pump-seed delay with piezo-driven delay-lines in the pump of last stage amplifiers (Pzt-DL on the right bottom of Fig. \ref{fig_synth_setup}).

\subsection{Basis Set for Stable and Controllable Waveform Synthesis}
\label{sec_CEP-RP}
\begin{figure}[!ht]
\centering
	\includegraphics[width=\textwidth]{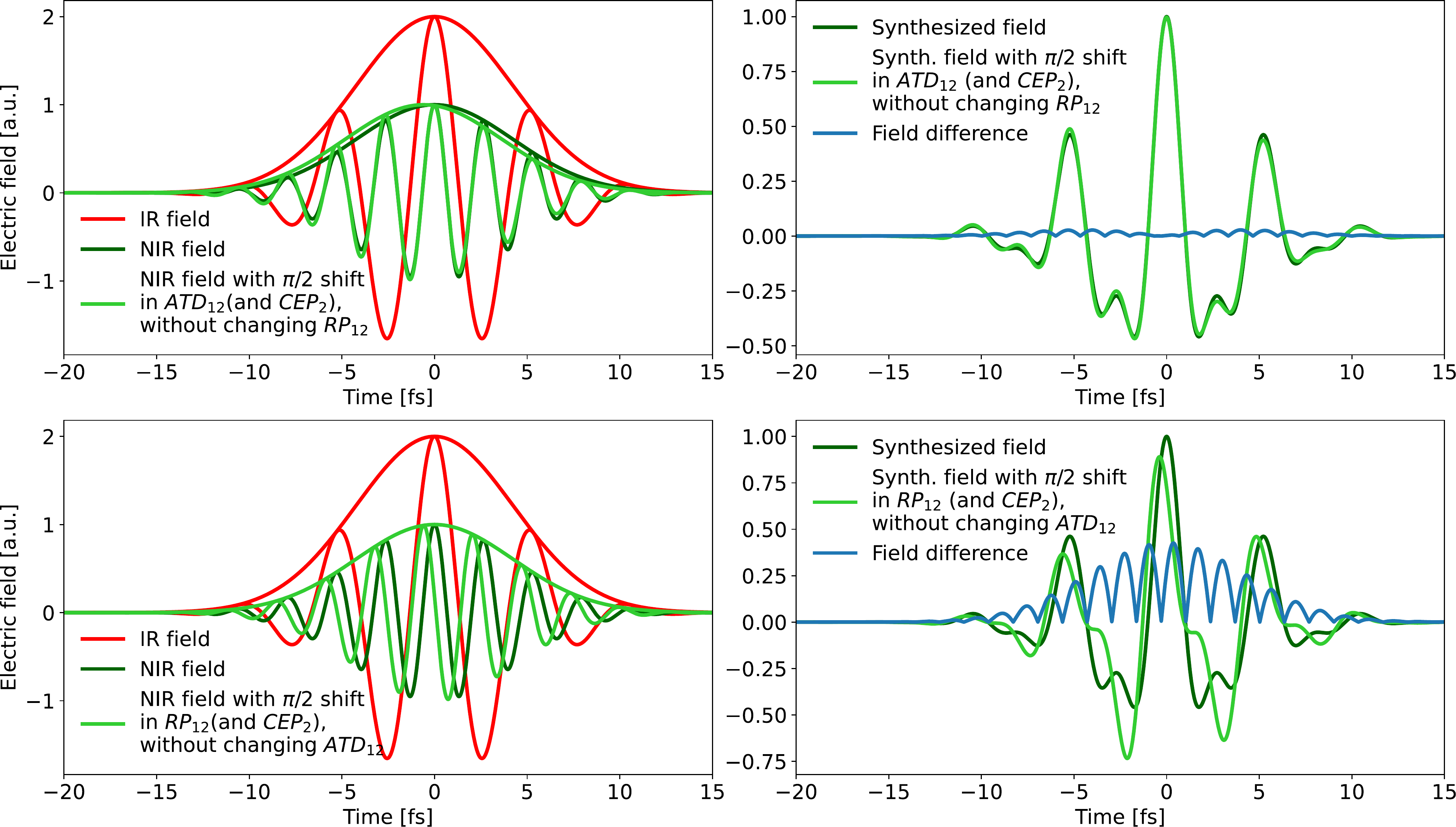} 
	\caption{\textbf{(a)} Electric fields of IR (red) pulse (1), NIR pulse (2) with $ATD_{12}=0$ (dark green) and NIR with $ATD_{12}=\pi/2$ (light green). \textbf{(b)} Synthesized fields with and without $ATD_{12}$ shift together with electric field difference (blue). \textbf{(c)} Electric fields of IR (red) pulse (1), NIR pulse (2) with $RP_{12}=0$ (dark green) and NIR pulse with $RP_{12}=\pi/2$ (light green). \textbf{(d)} Synthesized fields with and without $RP_{12}$ shift together with electric field difference (blue).}
	\label{synth_2ch_1+2}
\end{figure}

\begin{figure}[!ht]
\centering
	\includegraphics[width=\textwidth]{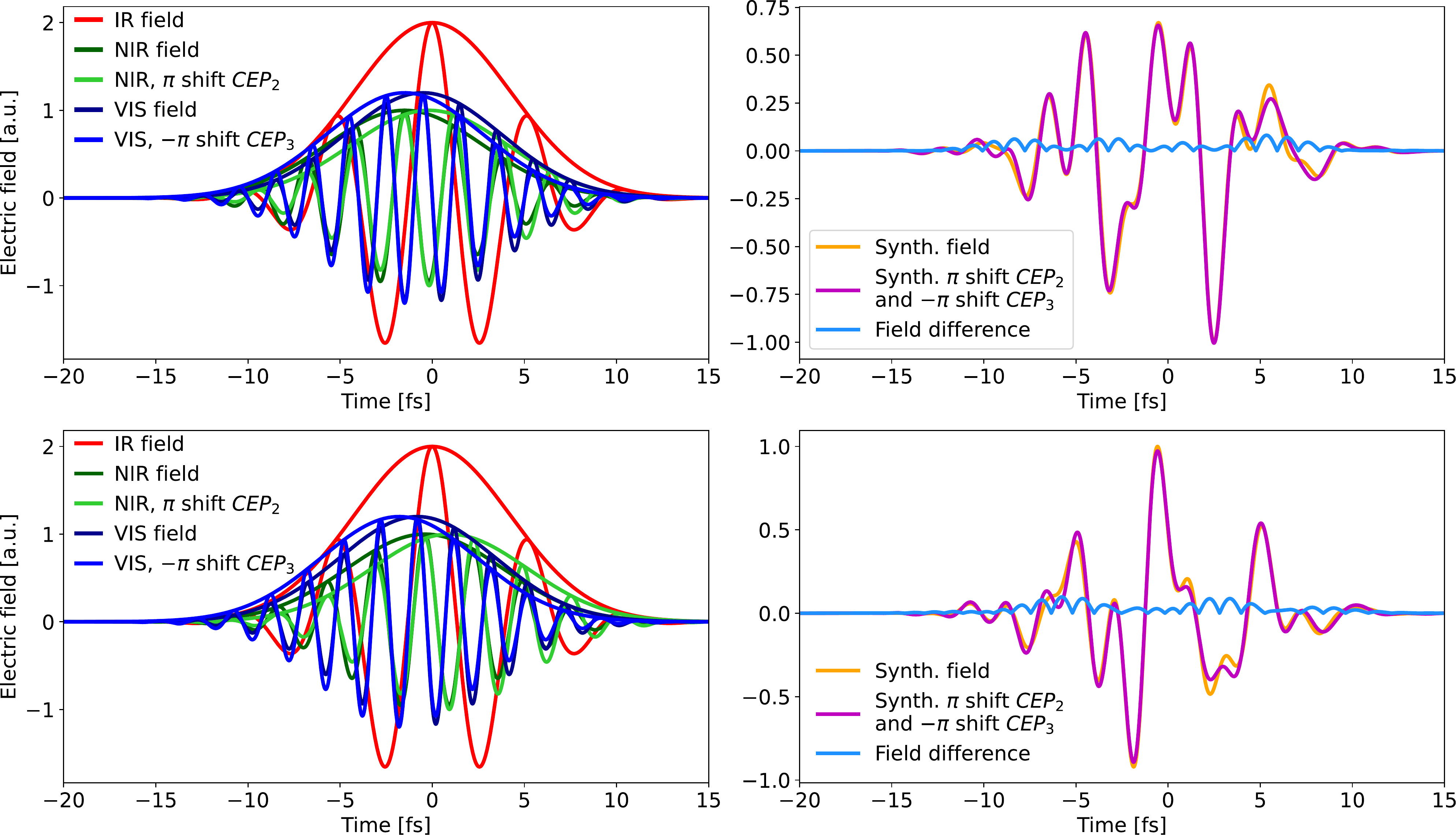} 
	\caption{\textbf{(a)} Electric fields of IR (red) pulse (1), NIR pulse (2) with $CEP_2 = 0$ (dark green) and with $CEP_2 = \pi$ (light green), VIS pulse (3) with $CEP_3 = 0$ (dark blue) and with $CEP_3 = \pi$ (light blue). \textbf{(b)} Corresponding synthesized fields with and without $CEP_{1, 2}$ shifts, together with electric field difference (blue). \textbf{(c)} Same as (a) for different $RP_{1, 2}$ setpoints. \textbf{(d)} Corresponding synthesized fields with and without $CEP_{1, 2}$ shifts, together with electric field difference (blue).}
	\label{synth_3ch_1+2}
\end{figure}

So far, we have analyzed the effect of timing variations on the CEP of the narrowband OPA seeder and on the broadband OPA amplifiers. In short, we concluded that: (i) the pump-seed delay in the seeder DFG stage can be efficiently exploited to control the CEP of the seeding pulses; (ii) the pump-seed delay jitter does not influence the CEP of broadband pulses significantly, emerging from the spectral channels of the PWS, at the compression point. The pulses from the spectral channels are then coherently combined to form the final waveform. We will now discuss the most effective way to control the synthesized waveform, that is which is the best set of waveform parameters and which is the most convenient way to observe and control them.\\
In a parallel parametric waveform synthesizer, the synthesized waveform is obtained by superimposing two or more pulses covering different bands of the optical region. In our PWS, the pulses to be synthesized cover 520-700 nm (VIS-channel, under development), 650-1000 nm (NIR-channel), and 1200-2200 nm (IR-channel). If we assume that (i) the pulses have a stable spectrum and energy (negligible shot-to-shot fluctuations) and that (ii) the pulses are a superposition of plane waves propagating along the same direction (meaning that here we neglect the transverse properties, which will be discussed in section \ref{sec_spatial_properties}), then the synthesized electromagnetic field waveform depends on: (i) the spectral phase of each pulse, (ii) the CEP of each pulse and (iii) the AT of each pulse. The complexity can be initially reduced by considering the spectral phase of each building pulse as fixed since most applications require them to be fully compressed. Given $N$ pulses to be synthesized (that is, a PWS with $N$ spectral channels), we would need to know $N$ CEPs and $N$ ATs. This adds up to $2N$ synthesis parameters in total. However, since a typical shift of arrival time of the $N$ pulses does not affect the synthesized waveform, we can consider one of the $N$ pulses as a reference and express the properties of the remaining $N-1$ pulses as differences with respect to the reference pulse. The reference pulse is called the \textit{master pulse}. This leads to $N-1$ ATDs with respect to the master pulse.\\
Let us now call $CEP_1$ the CEP of the master pulse. Similarly to what we did with the ATs, the CEPs of the $N-1$ pulses (the \textit{slave pulses}) can be equivalently expressed by relative phases ($RP_{12},...,RP_{1N}$) and ATDs with respect to the master pulse CEP and AT.
The synthesis parameters are now $2N-1$: $CEP_1$, $N-1$ RPs and $N-1$ ATDs.\\
Let us now assume to have full control over the CEP of the master pulse, that is $CEP_1$. We can now investigate the effects of ATDs and RPs on the synthesized waveform. To this end, we consider the synthesis of two pulses, pulse 1 being the master pulse and pulse 2 being a slave pulse. The behavior of the synthesized waveforms is shown in fig. \ref{synth_2ch_1+2}, where we set $CEP_1$ (IR channel) to a fixed arbitrary value and varied either $ATD_{12}$ or $RP_{12}$. In the first case (plots a,b) the $ATD_{12}$ changes by 0.67 fs, corresponding to $\pi/2$, while $RP_{12}$ stays fixed (therefore $CEP_2$ changed by $\pi/2$). In the other case (plots c,d) the $RP_{12}$ changes by 0.67 fs, corresponding to $\pi/2$, while $ATD_{12}$ stays fixed (therefore also in this case $CEP_2$ changed by $\pi/2$).
Small RP changes ($\Delta RP_{12} \ll 2\pi$) result in significant changes of the synthesized waveform. On the contrary, ATD variations that are small with respect to the duration of pulse 2 ($\Delta ATD_{12} << \tau_2$) do not influence significantly the synthesized waveform. This implies that a change in $CEP_2$ does influence the synthesized waveform only if accompanied by a change of $RP_{12}$. These simple observations suggest that the $RP$ is a stronger synthesis parameter with respect to the $CEP\&ATD$ pair. Therefore the $2N-1$ synthesis parameters can be reduced to just $N$ parameters, that is $CEP_1$ and $RP_{12},...,RP_{1N}$. This allows to tremendously simplify the waveforms control and stabilization system. This simplification entails a small decrease design freedom for the waveform, that however is irrelevant in the present state. To prove that, the maximum waveform difference attainable by controlling the additional degree of freedoms in a three channel synthesizer ($CEP_1 + RP_{12} + RP_{13}$ only vs. $CEP_1 + CEP_2 + CEP_3 + ATD_{12} + ATD_{13}$) is shown in fig. \ref{synth_3ch_1+2} in the limit case of $\Delta CEP_2 = \Delta CEP_3 = \pm\pi$ for two different arbitrary waveforms (different $RP_{12}$ and $RP_{13}$ setpoints).\\
In both cases, the additional freedom granted by $CEP_2$ and $CEP_3$ control allows only for small modifications to the synthesized waveform, showing a maximum field difference $<10\%$.
It is worth noticing that the waveform variations are very small also because we chose the IR channel, that contains most of the overall energy, to be the master channel. This is the optimal choice for this type of stabilisation scheme. If the energy of the different building pulses were similar, the variations on the synthesized waveform due to ATD fluctuation would be slightly larger, but still negligible in most applications.\\
The possibility of attaining extensive control over the synthesized waveform via $N$ parameters instead of $2N-1$ allows for a significant simplification of the active waveform control system, consisting of less detectors and fewer actuators. Moreover, it is also important to anticipate that the measurement noise associated with RP measurements is significantly lower than that associated with both ATD and CEP measurements since the latter usually require multiple nonlinear processes each (e.g., f-2f for CEP, a cross-correlation in the BOC for ATD), while the RP can be measured in a linear fashion (spectral interference) or at most via a single nonlinear broadening stage (e.g., mild spectral broadening). Jointly with a simpler measurement setup of synthesis parameters, the control of $CEP_1 + RP_{12}$ ensures superior stabilization performances respect to the control of $CEP_1 + CEP_2 + ATD_{12}$ in our current 2-channel system. This statement will be proven in later sections when we will look at active stabilization results (see \ref{Dual In-line Phase Meter} and \ref{sec_streaking}).\\
Now that we have defined the synthesis variables, that are $CEP_1$ and $RP_{12}$ ($RP_{13}$ will be added in future works with the VIS-channel), let us discuss how to control them. In section \ref{sec_pump-seed_narrowband} we already showed that the CEP of the idler pulses of the OPA-seeder can be efficiently controlled via a Pzt-DL in the beam-path of the pump (or of the seed). In principle, having the Pzt-DL on the signal path is better since, in this case, the CEP stabilization would not affect the ATD of the CEP-stable idler, whose temporal overlapped (after WL-generation) is necessary for the following OP(CP)As. However, this effect is minuscule since the Pzt-DL during stabilization, thanks to our compact OPA design, moves by just a couple of fs. According to eq. \ref{idler_CEP_narrowband_1}, a $2\pi$ CEP shift can be attained with a delay of $\approx 4.3$ fs, a small amount compared to $\tau_{pump} \approx 150$ fs. Practical considerations led to placing the Pzt-DL in the pump path of the second stage, as shown in the OPA scheme in fig. \ref{fig_CEP_top_beam}. Indeed, adding a delay-line on the signal path (between first and second stage OPA) would require adding more optics to the otherwise very compact Mach–Zehnder-type setup.\\ 
The idler pulses from the OPA-seeder drive the WLG stages that create the seeds for each spectral channel of the PWS. Since the WLG process is coherent (as quantified in tab. \ref{tab:WLG-phase-noise}), by acting on the idler's CEP, we can control the CEPs of the seed pulses. Moreover, in section \ref{sec_pump-seed_narrowband} we concluded that the CEP stability of the seeds is maintained during their amplification in the spectral channels without the need of active stabilization of the pump-seed ATDs. Therefore, by controlling the idler CEP in the OPA-seeder, it should be possible to control $CEP_1$, the CEP of the \textit{master channel output}. The stabilization and control of $RP_{12}$ can be obtained by using a Pzt-DL in the beam path of either the master or the slave pulse. In our PWS, we decided to use two delay lines instead of one. By doing so, we could use the first delay-line to stabilize fast $RP_{12}$ fluctuations with a short-range actuator, while the second delay-line is used to control $RP_{12}$ over an extensive range spanning hundreds of cycles. In order to achieve a high stabilization bandwidth, the first delay-line consists of a home-built short travel range piezo-driven actuator (identical to the one used in the OPA-seeder) actuating a 1-inch mirror that was placed on the signal path of the slave channel ($\sim0^{\circ}$ incidence) before the last stage amplifier (see Fig. \ref{fig_synth_setup}). Alternatively, this Pzt-DL could be placed after the last stage amplifier. These two options are equivalent in terms of waveform control, and the first was chosen for practical reasons. Instead, the second delay line is based on a commercial translation stage with position feedback (PI XXX), which allows for a long-range and nanometric precision, also beneficial to quickly scan or adjust the time-zero between the master and slave pulses.\\
One more Pzt-DL was added in the beam-path of the pump pulse of the last stage amplifier of each spectral channel (both master and slave). As discussed, the pump-seed delay in the spectral channels does not significantly influence the CEP of the amplified signal pulses, provided that these are used once temporally compressed. However, particular applications might require waveforms with specific chirp profiles \cite{Maas1998}. In this case, in order to provide excellent waveform stabilization, it might be necessary to additionally stabilize the pump-seed ATD in the last stage amplifiers (the booster stage) since the corresponding pump pulses undergo a completely separate beam path (see fig. \ref{fig_synth_setup}), leading to a few fs temporal jitter with respect to the signal pulses. Moreover, as outlined in fig. \ref{fig_pump_seed_spectra}, the spectrum of the signal pulses emerging from each spectral channel depends on the pump-seed delay; therefore, the possibility to remotely control it allows for optimizing the source easily.
As mentioned at the end of Sec. \ref{sec_CEP-RP}, these delay lines were recently used to stabilize the spectral shape of the OPA outputs, that can undergo small and slow (<1\,mHz) drifts due to thermalisation dynamics occurring in the external compressor (C2 in \ref{fig_synth_setup}) and influencing the pump-seed delay in the last OPA stages.\\
The control system designed to stabilize and shape the synthesized waveforms by driving the aforementioned Pzt-DLs will be presented in section \ref{sec_control_system}. In the final part of this section, we describe the multi-phase detector that allows measuring $CEP_1$ and $RP_{12}$.\\

\subsection{Dual In-line Phase Meter}\label{Dual In-line Phase Meter}
\begin{figure}[!ht]
\centering
	\includegraphics[width=0.75\textwidth]{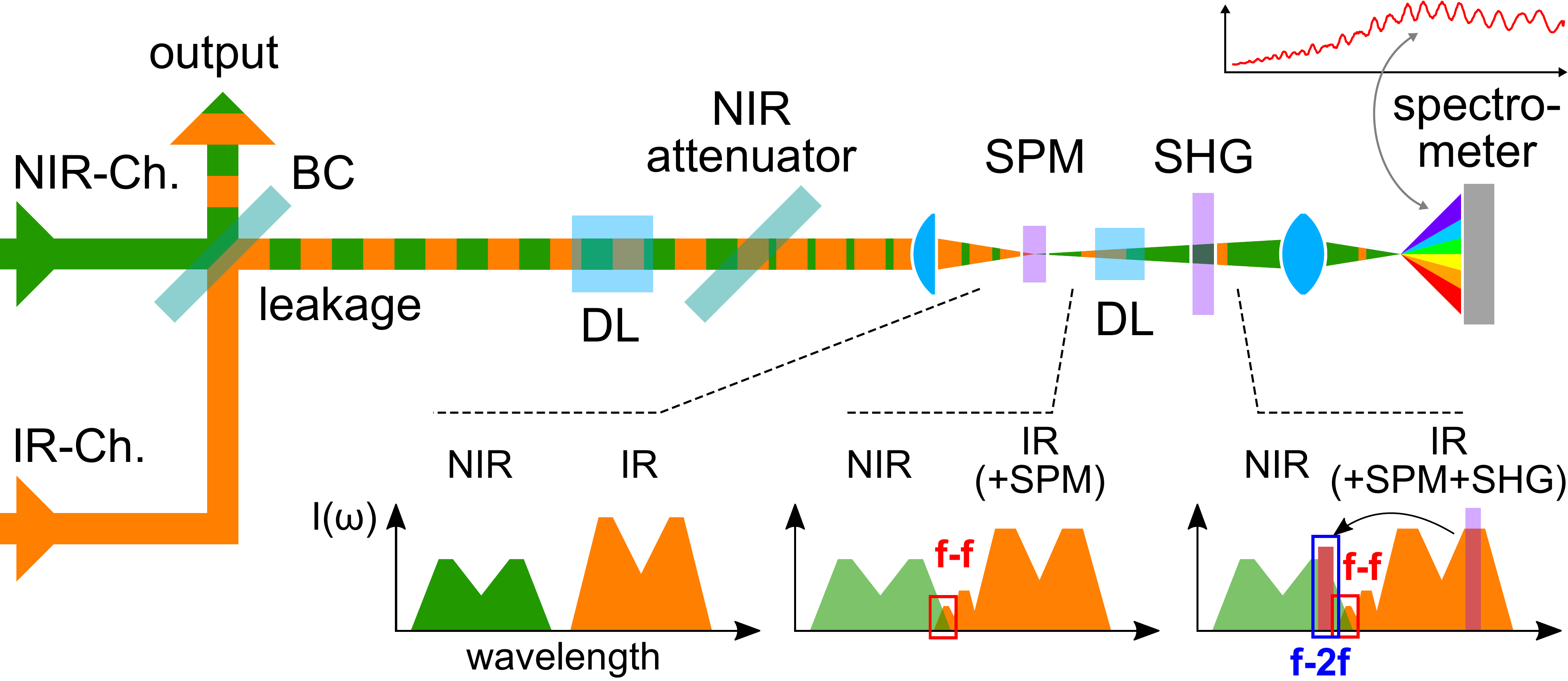} 
	\caption{Scheme of the dual-phase meter. The secondary output from the NIR/IR beam-combiner is used for CEP/RP detection. First the temporally overlapping NIR and IR beams are delayed (DL) by simple dispersion, then the NIR is attenuated before focusing into an SPM stage to predominantly broaden the IR-component. After a second delay element the beam is refocused into an SHG-stage for 2f generation for CEP detection before all beams are projected onto the same polarization plane and are observed by a single-shot spectrometer.}
	\label{fig_dual_phase_meter}
\end{figure}

\begin{figure}[!ht]
\centering
	\includegraphics[width=\textwidth]{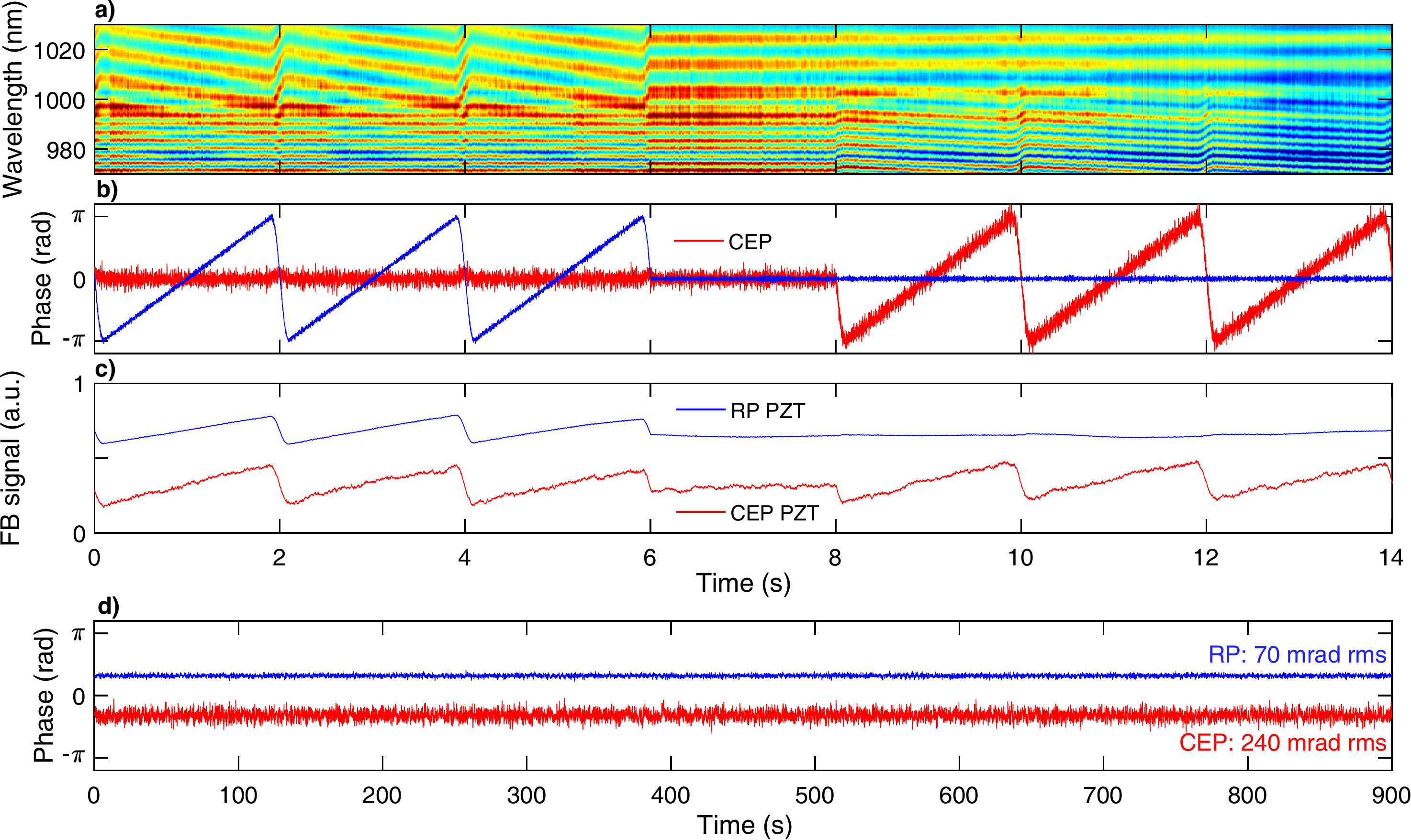}
	\caption{Dual Phase Meter Measurement: \textbf{(a)} Trace of observed spectral fringes with superimposed beats for CEP and RP detection. \textbf{(b)} Extracted CE-phase in red and RP values in blue. A locked triangular RP scan is performed (0-6\,s) and moved the both the RP and the CEP-actuator (in \textbf{(c)}) due to the CEP-RP convolution, in order to keep the CEP itself stable. A triangular locked CEP-scan (8-14\,s) only moves the CEP-actuator.
	\textbf{(d)}) Locked CEP and RP with remaining phase noise of 70\,mrad (RP) and 240\,mrad (CEP). Our control system stabilizes both phases and allows to manipulate the synthesis set points.}
	\label{fig_DualPhaseMeter_meas}
\end{figure}

In the previous chapters, the most effective synthesis parameters were defined ($RP_{12}$ and $CEP_{1}$), and suitable locations for corresponding actuators were identified. The measurement of synthesis parameters serves a twofold purpose: (I) stabilization of the multi-path interferometric setup within the PWS and (II) control of the final synthesized waveform, with resolution down to a fraction of the period of the optical cycles. The sampling of $RP_{12}$ and $CEP_{1}$ should preferably occur after the individual pulses have been recombined, so that no additional temporal jitter is introduced afterwards. For this purpose, the final beam combination optic(s) (see DM1 and DM2 in Fig. \ref{fig_synth_setup}) were designed to provide a replica at the secondary port with 5\,\% of the input pulse energy. To avoid possible drifts among the two synthesis parameter, a unique all-inline multi-phase meter capable of simultaneous detection of $RP_{12}$ and $CEP_{1}$ was developed \cite{Rossi_UFO_2019} (see Fig. \ref{fig_dual_phase_meter}).
The basis of this phase meter is spectral interferometry, where spectral beats need to be created by nonlinear conversion(s) to retrieve the CE and relative phase. In our current case, the spectrally non-overlapping NIR and IR-channel pulses require a linear spectral beat ($f_1$-$f_2$) in order to retrieve the relative phase $RP_{12}$. For this purpose, a mild SPM-based broadening in bulk is exploited to spectrally broaden the short-wavelength wings of the IR-channel pulse and create the desired spectral overlap with the NIR-channel pulse around 950\,nm. To selectively apply the SPM-broadening to the IR pulse only and avoid other RP-dependent cross-sensitivities (such as cross-phase modulation) that might appear in such an in-line scheme, the NIR-channel pulse is attenuated and delayed before both pulses are focused into a few mm of YAG acting as SPM-stage. Refocusing the beam in a subsequent SHG-stage additionally creates an ($f_2$-$2f_1$) beat between the frequency-doubled long-wavelength leg of the IR-channel pulse and the NIR-channel pulse, which is proportional to both $RP_{12}$ and $CEP_{1}$. The SHG process produces not only this inter-pulse $f_2$-$2f_1$ beating but as well an intra-pulse $f_1$-$2f_1$ beating. Such inter-pulse CEP-detection brings the disadvantage of being sensitive to $RP_{12}$ too; on the other hand, it offers a higher signal-to-noise ratio with respect to the intra-pulse CEP detection. Moreover, since the RP can be stabilized with really low residual noise, this does not negatively impact the CEP stabilization. The delays among the replicas of NIR pulse, IR pulse and IR-SH pulse used in the multi-phase meter correspond to the spectral beat frequencies detected on the spectrometer. In order to have the beat frequencies well separated and not too close to zero-frequency, the delays can easily be adjusted by inserting suitable dispersive materials in between the nonlinear optical components or by exploiting different polarization states in combination with birefringent materials. Finally, all signals are projected onto a common polarization plane with a wire-grid polarizer and observed by a custom-made FPGA-based single-shot spectrometer (see Chapter \ref{sec_control_system}). The spectral regions in which the different beating signals are observed can be superimposed (e.g. $SHG_{IR}$, $SPM_{IR}$, NIR-channel) and can be observed by the same spectrometer in order to use the optical energy efficiently. The only requirement to retrieve the corresponding phases is that these spectral beats appear at different beat-frequencies. A Fourier-transform of the observed optical spectrum exhibits magnitude signals at those beat frequencies. Tracking and unwrapping the phases of those Fourier-components allows to isolate the desired signals associated with synthesis parameters, which then can be used to generate the error signals for the active stabilization system of the PWS, as shown in Fig. \ref{fig_DualPhaseMeter_meas}).
We observe a low phase noise for the RP measurement, which involves SPM only, and higher noise for the CEP measurement, which additionally requires SHG that introduces substantial intensity-to-phase coupling. While the WLG-based RP measurement can yield as low as 30\,mrad rms of measurement noise (see Tab. \ref{tab:WLG-phase-noise}), the SHG for CEP detection adds 150-200 mrad of measurement noise on top of the noise of the source.

\FloatBarrier
\section{Technical Implementation}\label{sec_technical_implementation}
One of the most significant challenges to achieve stable synthesis with a parallel synthesizer scheme is realizing an optomechanical setup with sufficiently high stability and implementing an active timing stabilization with high bandwidth and reliability. While the nonlinear optics techniques to achieve and maintain phase-stable pulses and measure them were discussed previously, now a focus will be on the low-latency control system and the optomechanical implementation of the PWS setup.\\
No matter what the capabilities of the active stabilization system are, the optical setup should exhibit the lowest possible phase drifts already in passive operation. This goal can be achieved partly by selecting the correct means of pulse generation, amplification and phase detection methods as previously laid out. Additionally, the interferometric setup needs to be realized with high precaution to avoid long-term drifts or (undamped) vibrations of the optical elements, affecting the phase stability and the beam pointing.\\

\subsection{Active Phase Stabilization and Control System}
\label{sec_control_system}
\begin{figure}[!ht]
\centering
	\includegraphics[width=\textwidth]{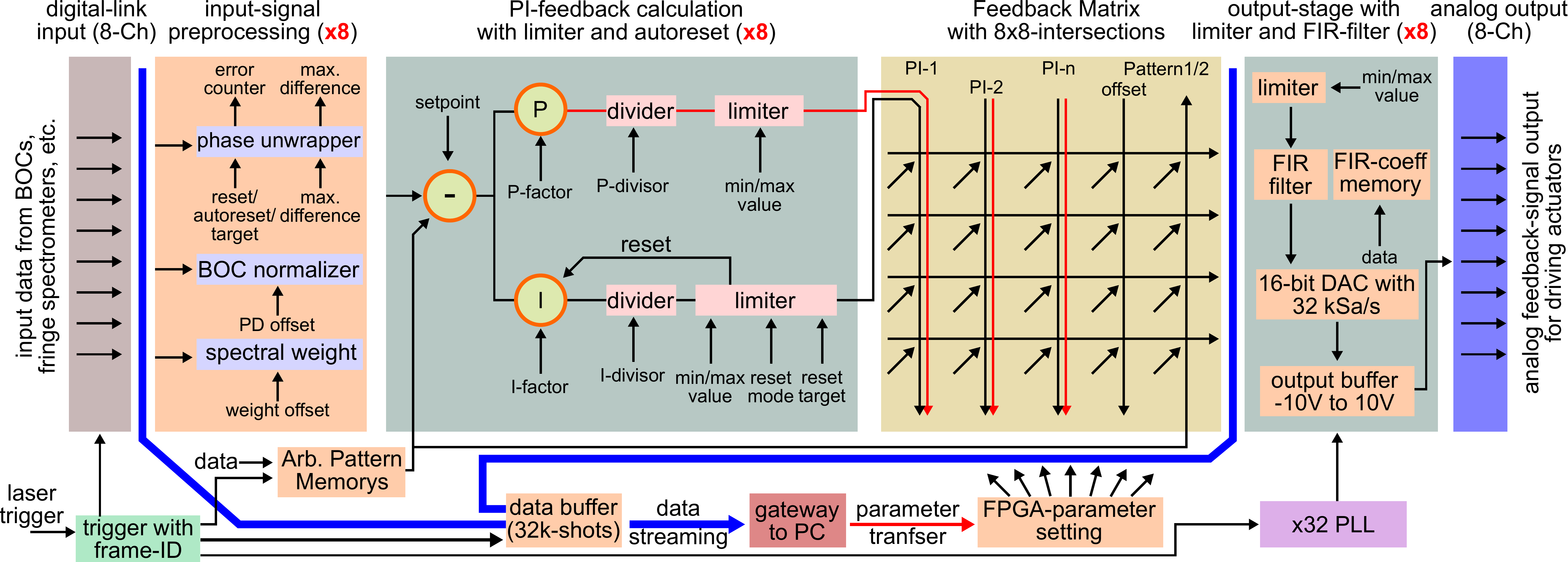} 
	\caption{Overview of the FPGA-based feedback system. Incoming timing parameters are preprocessed such as unwrapping for phase-values or normalized balancing for BOC-inputs. Then the processed observables enter a PI-controller section before a matrix allows to (de)couple the individual feed-backs and derive a suitable control signal to the multiple timing actuators (fast and slow). With this system, the laser pulses are indexed and all input/output parameters are recorded to maintain a full insight on the synthesizer state at any given time.}
	\label{fig_FPGA_signal_path}
\end{figure}

\begin{figure}[!ht]
\centering
	\includegraphics[width=\textwidth]{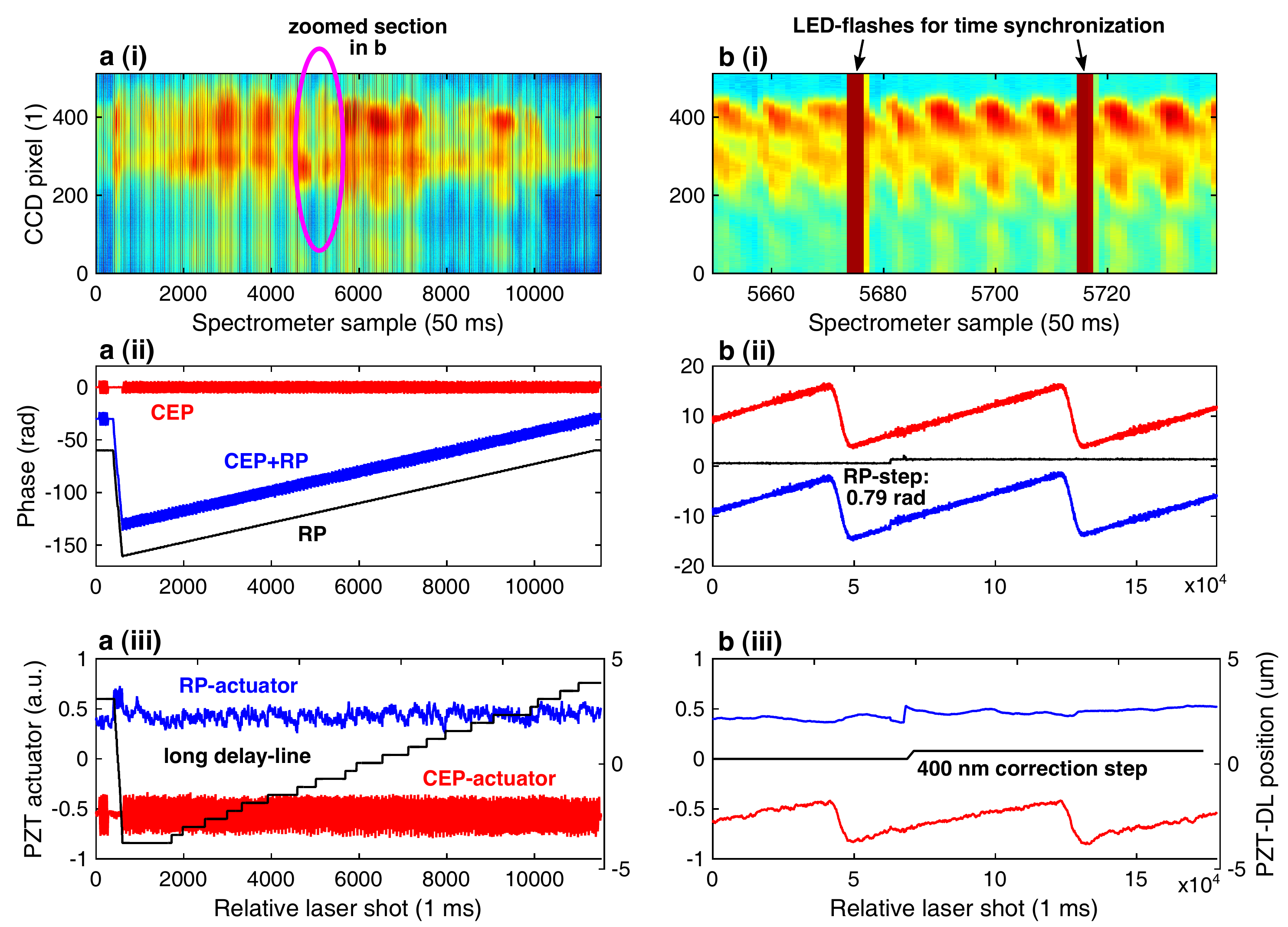} 
	\caption{Exemplary trace of a synthesized waveform scan with measured experimental observable in \textbf{(a)} and zoomed-in section in \textbf{(b)} for one scan-cycle. The observable \textbf{(i)} are here raw HHG-spectra (50\,ms integration time). \textbf{(ii)} Measured phase observables where the RP follows a plain ramp and CEP is driven as a saw-tooth modulation for a full 2D parameter mapping. \textbf{(iii)} Corresponding actuator signals for the CEP-actuator (red) RP-actuator (blue) and a long-range stage with correction steps (black), to keep the short-range actuators in their dynamic range.}
	\label{fig_phase_lock}
\end{figure}

The basis of the optical means to gain access to the relevant timing parameters is discussed with the introduction of the multi-phase meter for detecting the CEP and RP values. Phase measurement techniques are favored over other envelope timing tools such as balanced optical cross-correlators (BOC \cite{Schibli_OptLett_2003}) for our application since phase changes have a much bigger impact on the synthesized waveform compared to envelope arrival time changes. Moreover, the in-line phase measurement is less prone to thermal drifts and more robust with respect to pointing changes and beam profile changes. On the other hand, the BOC can measure an absolute timing relation between two pulse envelopes, while a phase measurement can only determine relative phase changes with respect to the previous phase value. If a phase difference greater than $\pi$ occurs between two consecutive laser shots (\textit{phase-jumps}), the relative change will be ambiguous and the lock will be void. In this case, the synthesized waveform changes in an unknown direction and, if a waveform scan is being performed, it can no longer be brought into relation to previously synthesized waveforms. This ability is crucial during a full CEP-RP scan, for example, aiming at studying the generated HH-emission \cite{Yang_NatComm_2021}, or when keeping a fixed set-point during an attosecond streaking measurement.\\
This circumstance leads to the conclusion that a scheme based on phase measurements practically requires a pulse repetition rate of >100\,Hz and, if possible, a single-shot and every-shot evaluation of the spectral interference in the phase-meter. Out of these demands, we developed a dedicated spectrometer based on an FPGA-based circuit that processes the data of a linear image sensor (Hamamatsu, S10453 (<1.1\,$\upmu$m), G9208-256W (1.2-2.3\,$\upmu$m)) within our spectrometer with low latency and on an every-shot basis at 1\,kHz rep.-rate. The FPGA computes the necessary Fourier-transforms and determines the unwrapped phase information within a few hundred of $\upmu s$ from the recorded spectra to provide sufficient time for the feedback system to move the piezo-driven delay-lines accordingly to compensate for the observed drifts with only 1-2 laser shots of latency. Besides these single-shot phase tracking spectrometers, an FPGA-based feedback system was implemented to calculate the individual feedback signals to actuate the corresponding piezo-driven delay lines (see Fig. \ref{fig_FPGA_signal_path}). This control system features phase-unwrappers with error detection, normalization calculation (for BOCs), FIR-filters, a feedback matrix for orthogonalization, range-limiters and arbitrary waveform generators.
For actuation, we implemented in the design fast acting (400\,Hz bandwidth) and low range (3\,$\upmu$m) piezo-driven ring actuator (Noliac, NAC2125-A01) moving 1-inch mirrors. In the CEP-stable front-end we can control the overall CEP over a few cycles, which is sufficiently wide due to its 2$\pi$ periodicity. An additional ring-piezo actuator is placed before the 3rd stage OPA of the NIR-channel to control the RP. Additionally, we use a 25\,mm long-range stick-slip stage (Physik-Instrumente, N-565.260) to adjust the rough phase-delay, respectively the temporal separation between the NIR and IR pulses. The feedback system allows configuring the phase set-points, the proportional and integral feedback parameters and the coupling coefficients. Via scripts executed on a control computer, we can also perform parameter scans of the waveform with different patterns, such as a sinusoidal, a linear ramp or a step function (see Fig. \ref{fig_phase_lock}).
We obtain a data-stream from the control system containing all input and output parameters for each indexed laser pulse so that other data recorded in parallel (e.g. HHG-spectra) can be fully synchronized with corresponding RP-CEP values in post-processing. In this mode, highly efficient and flexible data acquisition is achieved since the additional experimental data can be collected without dead-time.\\
A complete reconstruction of the synthesized waveform can be recorded by an attosecond streaking trace, allowing to determine numerically the absolute value of the RP between the individual sub-pulses and their CEPs. By splitting the spectrum of the waveform obtained from the streaking trace and normalizing its spectral intensity to the one measured with an optical spectrometer, it is possible to obtain a quite accurate reconstruction of the electric field of each pulse. The spectral phase can then be compared with the 2DSI data to check for consistency. After this procedure, if the synthesis parameters are scanned, the corresponding waveforms can be derived numerically by applying the relevant CEP and RP offsets.

\subsection{Opto-mechanical Setup}
\label{sec_Optomechanical_design}
\begin{figure}[h!]
\centering
	\includegraphics[width=\textwidth]{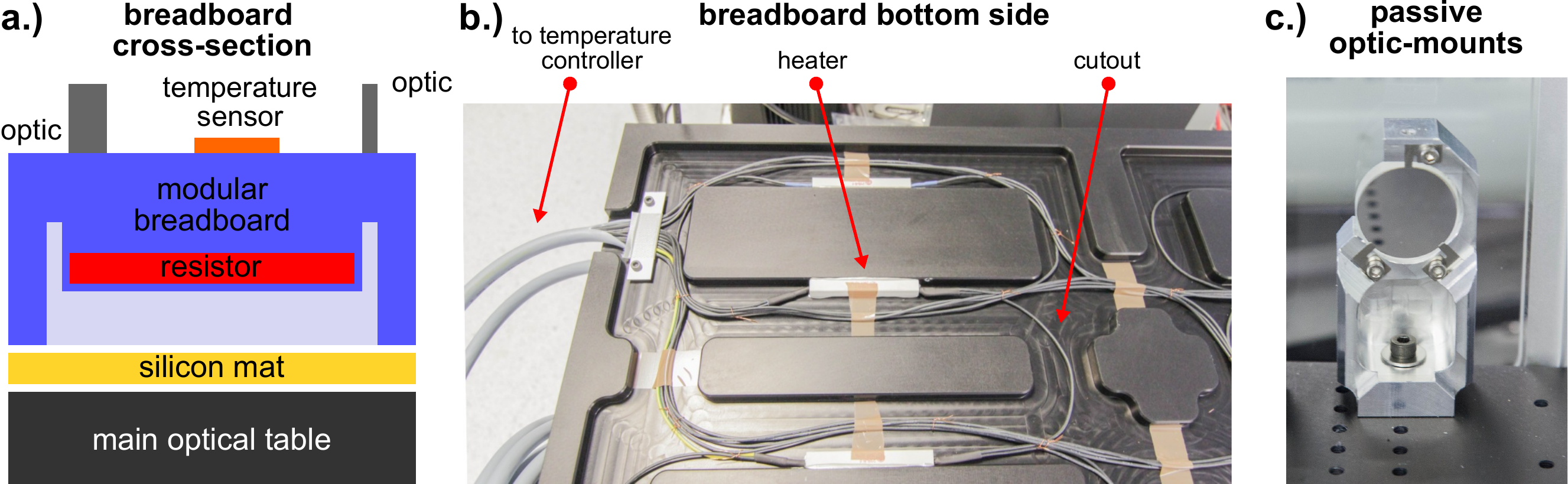} 
	\caption{Details on the opto-mechanical implementation: \textbf{(a)} Layering of the breadboard modules. \textbf{(b)} View of the bottom of the breadboards with vibration free heating elements for multi-zone temperature stabilization. \textbf{(c)} Highly stable custom-made optic mounts holding the optic either via springs or via UV-cured glue.}
	\label{fig_breadboard_setup}
\end{figure}

The passive stability of the optical setup is of paramount importance to achieve stable synthesis. In order to minimize phase and pointing fluctuations, the number of reflections and the beampath of the setup need to be minimized. Due to the complexity of the PWS setup, we opted for a modular design, where each module was individually optimized. The different modules are: the CEP-controlled seeder, the dual-beam delay lines (next to the seeding FE in Fig. \ref{fig_synth_setup}), the spectral channels and the multi-phase meter. Each module is implemented on an individual 5 cm thick custom-made aluminum breadboard resting on a 4\,mm thick silicon-rubber mat placed directly on the optical table. The silicon rubber helps to even out the weight distribution and avoid bistability (see Fig. \ref{fig_breadboard_setup}). Air fluctuations are effectively prevented by enclosing each module. While the seeder module is fully sealed by using AR-coated windows for the beams to propagate in or out, the broadband spectral channels have empty holes at the beam output to avoid additional dispersion and possible nonlinearities. Most of the optomechanics are also custom-designed and CNC-milled out of aluminum (see Fig. \ref{fig_breadboard_setup}c). On the one hand, this allows achieving a $\sim3$ times higher density of optical components with respect to conventional optomechanics, granting the possibility to significantly shorten the beampath and the footprint of the setup. On the other hand, the custom components were designed to have only the necessary degrees of freedom (most mirrors do not have any adjustment screw), enabling significantly higher passive stability which simplifies the alignment procedure. The optics are fixed by spring-loaded levers or UV-cured glue instead of a regular top-screwing mechanism to avoid stresses that would deteriorate the phase front of the pulse. Another critical aspect of the optomechanical setup is temperature stabilization. Temperature fluctuations of the environment lead to deformation of the optomechanics resulting in pointing and beampath length changes. We observed that these changes are reversible for minor air temperature variations (<1 K) but non-reversible when large temperature changes persist for tens of minutes or longer. To decouple the PWS setup from environmental influences, an active temperature stabilization system is implemented in each breadboard. Unlike laser amplifiers, where a significant amount of heat has to be removed due to the quantuum-defect, the modules of the PWS are OPA-based and thus without any significant heat load. This allowed us to avoid water-cooled breadboards that would require noisy chillers and pumps that could potentially introduce vibrations due to the water flow and decrease the overall reliability of the setup. Instead, we use electric heaters to temperature stabilize the breadboards of each module. 
On the surface of each breadboard 8 high precision sensors (Pt-1000) measure the temperature of different zones. On the bottom of the breadboard, 16 resistors, grouped by 2 and placed in specific cut-outs, are used to stabilize the temperature of each zone individually by having a set-point 2-3 K above room temperature. The set-point is chosen to be slightly above the maximum temperature we usually experience in the laboratory. Thanks to these temperature-stabilized breadboards, our seeder achieves an out-of-loop temperature stability of 3.3 mK rms over the course of several months, eliminating this environmental influence and the associated misalignment.

\section{Spatial Properties and Beam Combination}\label{sec_spatial_properties}
\begin{figure}[h!]
\centering
	\includegraphics[width=0.7\textwidth]{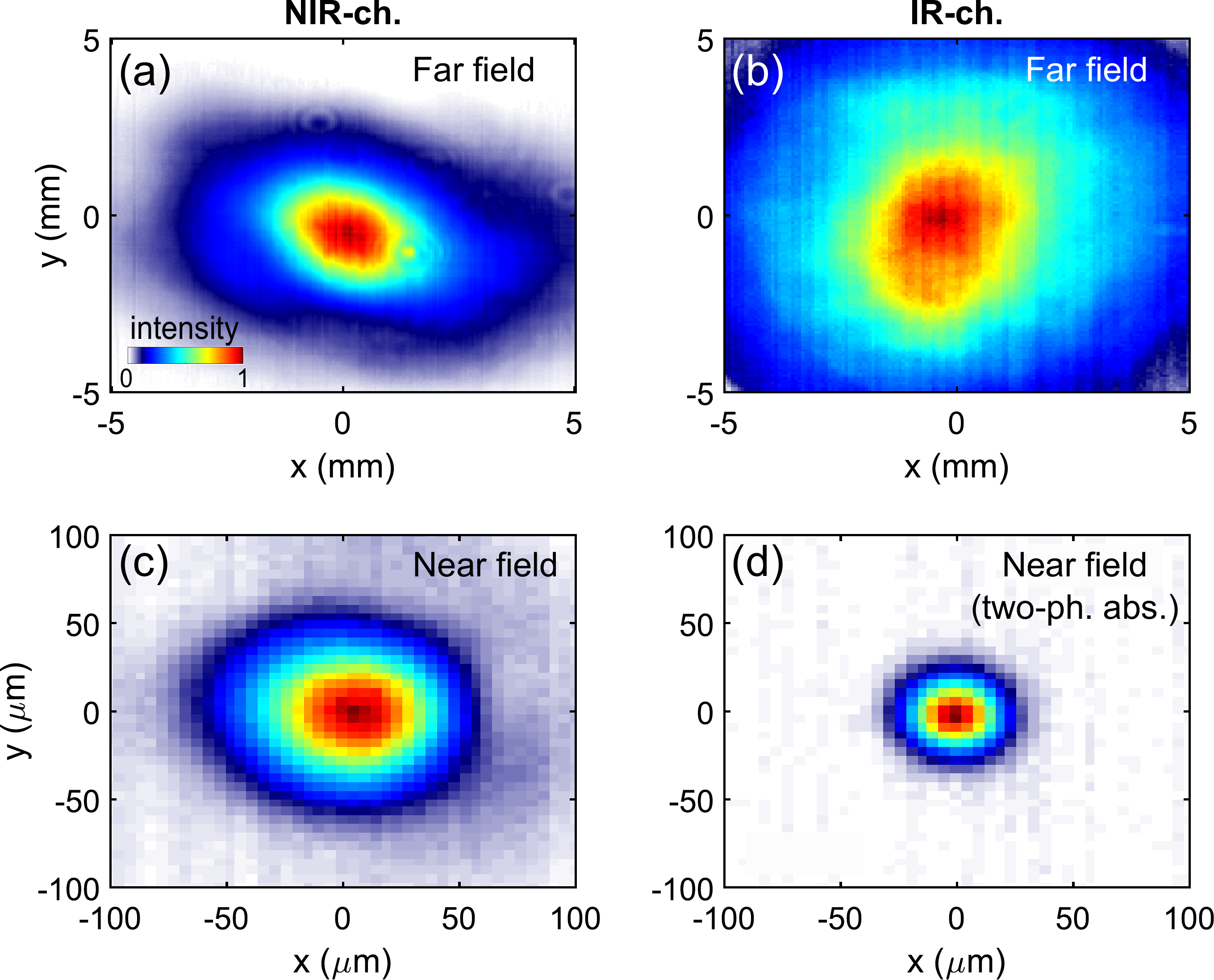} 
	\caption{NIR and IR channel far- and near-field beam profiles. Far-field beam profiles of \textbf{(a)} NIR and \textbf{(b)} IR channels and their respective near field \textbf{(c-d)} after a spherical mirror ($f = 500$ mm). A Si-based detector was used for (a), (c-d) and a pyroelectric array detector for (b). \textbf{(d)} shows the two-photon absorption signal resulting from the focused IR channel.}
	\label{beamcharac_fig1}
\end{figure}

\begin{figure}[h!]
\centering
	\includegraphics[width=0.7\textwidth]{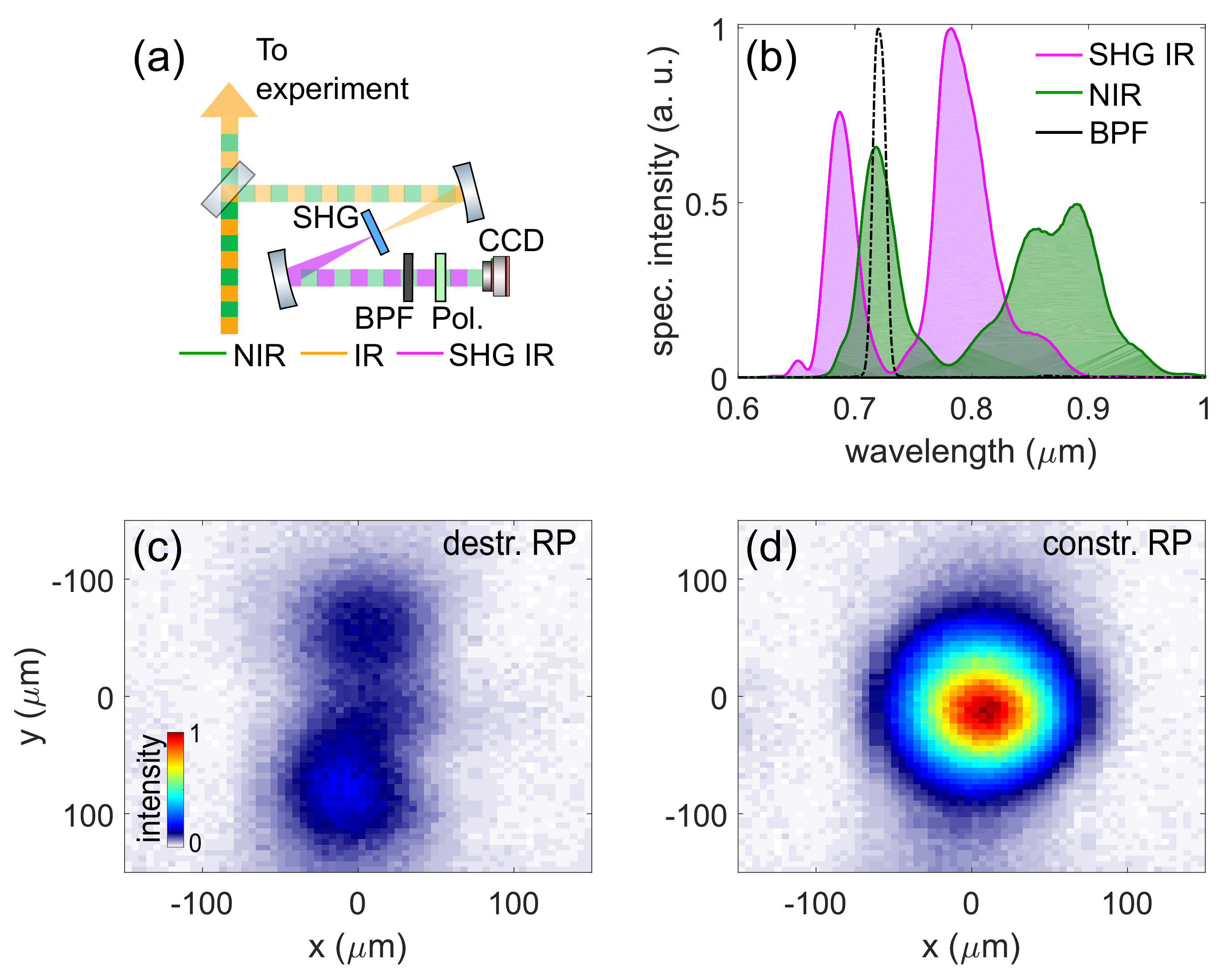} 
	\caption{Spatiotemporal overlap of NIR and IR beams. \textbf{(a)} Experimental setup. A pick-up deviates a portion of the beam towards the beam analysis setup. Second harmonic generation (SHG) of the IR beam is produced to beat with the NIR beam. A  band-pass filter (BPF) ensures a visible fringe contrast, and a polarizer (Pol.) allows to equalize the contribution of NIR and SHG of IR beams. A CCD detector records the resulting spatial interference. \textbf{(b)} Corresponding spectra. Spatiotemporal superposition between both beams at different RP values showing destructive \textbf{(c)} and constructive \textbf{(d)} interference.}
	\label{beamcharac_fig2}
\end{figure}

Pulse synthesis needs not only the temporal superposition between constituent pulses but also their spatial overlap. Moreover, the full spatiotemporal overlap requires shot-to-shot and long-term stability to guarantee highly-reproducible synthesized waveforms. 
The spatial overlap of all constituent pulses requires characterizing each beam profile evolution along the propagation axis. This characterization not only allows evaluating the beam quality but also to assure that, after a common focusing element, each pulse reaches its focus at the same longitudinal position. By doing so, it is possible to estimate the corresponding intensities and the intensity ratios between the pulses of each channel at the interaction point (usually close to the focus). The latter particularly impacts the shape of the synthesized field, therefore playing a significant role in nonlinear light-matter interaction. Moreover, phase-matching nonlinear processes, such as HHG, with a focused single-color beam is highly dependent on the beam waist at the focus, and its corresponding Rayleigh length \cite{Gaarde_2008}. Consequently, combining constituent beams with similar Rayleigh lengths (different focus sizes) or similar focus sizes (different Rayleigh lengths) can strongly affect phase-matching. The reason is that, in each case, the intensity ratio and relative phase between the beams evolve differently across the focus. In the current PWS, the beam sizes at the focus were matched, meaning that the far-field beam waist ratio between IR and NIR channels is $\approx 2$.
 The spatial characterization of the PWS is challenging, as the output covers 1.7 octaves of spectral bandwidth, and beam cameras capable of capturing the entire bandwidth are currently rather expensive. In addition, for many nonlinear, strong-field experiments, tens of micrometer of focus diameter are typically necessary, and pixel sizes able to resolve such beam dimensions are unfortunately only standard in silicon (Si)-based detectors, which are spectrally sensitive up to $\approx 1 \upmu m$. Fig.\ref{beamcharac_fig1} (a-b) shows the far-field NIR and IR beam profiles after beam combination. The IR channel beam profile was measured using a pyroelectric array detector (Spiricon, Pyrocam III-HR). Fig. \ref{beamcharac_fig1} (c-d) shows the focused, near-field beam profiles after focusing with a spherical mirror ($f = 500$ mm). As the resolution of the pyroelectric array detector is insufficient to resolve the focused IR beam, the Si-based detector measured the corresponding two-photon absorption signal (see Fig. \ref{beamcharac_fig1} (d)). Because such a nonlinear process scales quadratically with the IR beam intensity, the factor $\sqrt2$ must multiply the measured beam size. This method was cross-checked by knife-edge measurements, which differed by < 2\%. The resulting beam diameters at focus were measured to be $\sim$ 120 and $\sim$ 112\,$\upmu$m for the NIR and IR channels, respectively, with a longitudinal focus position difference of $\approx 100$ $\upmu m$.
To check whether the beams are overlapping in space and time at 1 kHz repetition rate, hence for every single shot, we built a characterization station that evaluates the spatiotemporal overlap of both beams of each channel in the near-field (see Fig. \ref{beamcharac_fig2} (a)). From a weak replica of the combined beam, the IR beam is frequency-doubled in a 100 $\upmu$m-thick Type-I BBO-crystal (see Fig. \ref{beamcharac_fig2} (b)) and is spatially-interfered with the NIR beam. A band-pass filter (BPF) enhances the fringe contrast. After that, a wire grid polarizer is placed just before a high-speed camera (Basler acA640-750um) to equalize the contribution of NIR and IR-SHG. By operating in this every-single-shot mode, the spatial interference does not average out, and its evolution can be directly linked to CEP and RP variations. Figure \ref{beamcharac_fig2} (c-d) exhibits the corresponding near-field spatial interference when the beams are spatially overlapped, with the RP set for (c) destructive and (d) constructive interference. This characterization is realized in parallel to an ongoing experiment, thus permitting continuous, online monitoring of the spatiotemporal overlap.
For the long-term stabilization of the beam pointing, an active stabilization system (TEM, Aligna) is employed for each constituent beam. After beam combination, a reflection from a thin glass plate derives a weak replica of the synthesized beam, which is then re-split into NIR and IR beams by dichroic mirrors and fed to the position-sensitive detectors (PSDs). The IR beam is additionally frequency-doubled in front of the PSD to match the spectral sensitivity of the Si-based detector.

\section{Waveform Reproducibility Characterized by Attosecond Streaking}\label{sec_streaking}
The measurement of the precise synthesized waveform is quite challenging for most pulse characterization methods, mainly due to the broad bandwidth. Apart from temporal pulse characterization via 2DSI, attosecond streaking was employed, which directly maps the vector potential of the electric field in an oscilloscope-like trace. In this technique, the generated isolated attosecond XUV/soft X-ray pulse ionizes a gas, and an optical field, in this case a replica of the synthesized pulse, modulates the kinetic energy of the liberated electrons. Due to the single-step photoionization, the photoelectron spectrum closely resembles the spectrum of the attosecond pulse, lowered by the ionization potential of the gas. The replica of the optical, synthesized field, whose polarization is oriented parallel to the drift tube of the time-of-flight electron spectrometer, streaks the electron energies. The energy modulation depends on the delay between the ionization time (by the attosecond pulse) and the optical field. By sequentially recording photoelectron spectra at different delays between the isolated attosecond pulse (IAP) and the optical field, it is possible to access both the vector potential of the optical waveform and (by retrieval/reconstruction algorithms) both the optical electric field and spectral phase of the attosecond pulse. Due to the time-consuming delay scan and the highly nonlinear processes involved, the attosecond streaking technique is among the most demanding pulse characterization measurements in terms of shot-to-shot stability of the pulses involved, representing the ultimate benchmark for the PWS.
\begin{figure}[h!]
\centering
	\includegraphics[width=\textwidth]{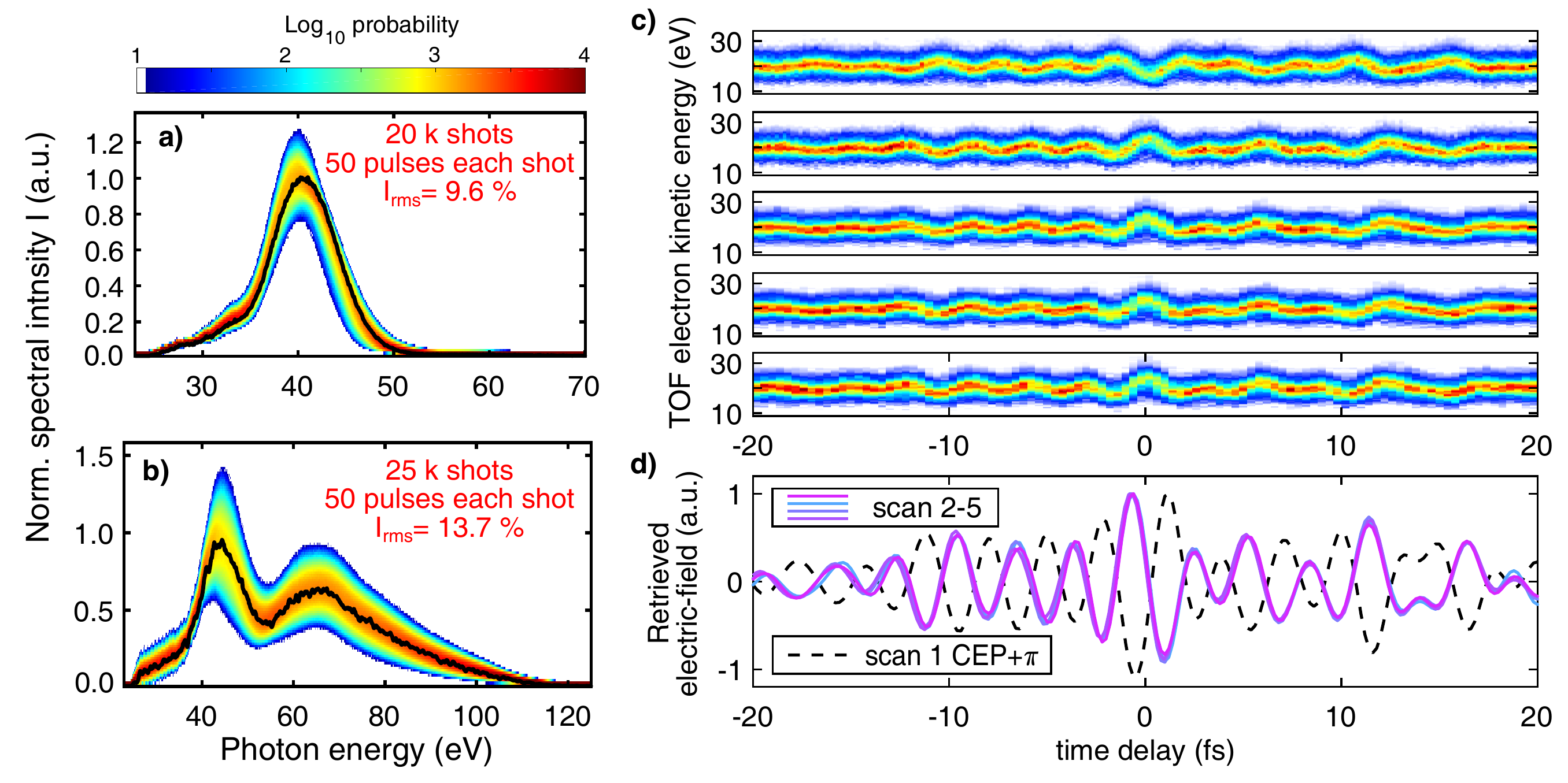} 
	\caption{\textbf{(a, b)} Spectral stability of isolated attosecond pulses generated via HHG with two different pulse synthesis settings as logarithmic color heat map and the mean spectrum over 20, and 25k shots each (black). The intensity of the generated attosecond pulses reach a high stability with a remaining rms fluctuation between 10-14\%. \textbf{(c)} Top: Two finely sampled attosecond streaking traces with 180$^\circ$ CEP shift. Bottom three repeated streaking traces with the same CEP-setting demonstrating repeatability. \textbf{(d)} Retrieved electric-fields from \textbf{(c)} by a center of mass algorithm \cite{SilvaToledoMA_HBSLIC2020}.}
	\label{waveform_stability}
\end{figure}
The waveform must be highly repeatable to generate identical isolated attosecond pulses in the high-harmonics gas source. In Fig. \ref{waveform_stability}a, we show the spectral intensity stability of different IAPs generated by the PWS. We observed an rms stability of 7\% at  40\,eV when averaged over 50 shots. Multiple consecutive attosecond streaking traces were recorded with identical waveforms over 2 hours (4 successive traces) in order to evaluate the long-term waveform stability (See Fig. \ref{waveform_stability}b). A center-of-mass waveform retrieval algorithm yields retrieved electrical waveforms with an excellent agreement and without significant drifts or waveform modifications (see Fig. \ref{waveform_stability}c). Observed differences in the retrieved fields can be attributed to interferometric drifts between the IAP and the streaking field, as the attosecond beamline was not actively stabilized. Additionally, to demonstrate control over the synthesized waveform, a $\pi$ shift in the CEP was introduced in one streaked waveform \cite{SilvaToledoMA_HBSLIC2020}. 

\FloatBarrier
\section{Conclusions and Perspectives}\label{Conclusions}
This paper presents the enabling techniques for parallel parametric waveform synthesis (PWS). With these technologies, the PWS can deliver stable and controllable non-sinusoidal optical waveforms with sub-cycle durations, mJ-level energy at 1 kHz repetition rate. Since this technology is based on OP(CP)A, it is intrinsically scalable to higher energy and power. We modeled the phase propagation in the different OP(CP)As of the PWS. The contribution of pump-seed temporal jitter in the narrowband OPA-seeder was quantified in a previous publication, and the conditions to optimize the CEP stability were obtained. Here we extend the analytic description to the broadband OP(CP)As of the spectral channels proving that small pump-seed relative arrival time fluctuations do not influence signal CEP if large stretching ratios are used, and the pulses are fully recompressed at the interaction point. These considerations on the timing dynamics clear the way to energy and power scaling of PWS since different laser technologies could be adapted.\\
So far, we developed a two-channel (NIR and IR) OPA-based PWS that delivers a 1.7-octave bandwidth, FWHM durations down to about 0.6 cycles, and energies up to $\sim 0.5$\,mJ. The system is ready to be upgraded by an additional channel covering the VIS region.\\
The realization of PWS setups requires a careful choice of optical, mechanical, and active stabilization techniques. CEP-controlled seed drivers are derived from non-CEP-stabilized laser beams via DFG in a two-stage OPA. Multi-octave wide, yet highly phase coherent, seed pulses can be obtained by spectral broadening of CEP-controlled driver pulses in separate WLG-stages, each optimized for the phase-matching bandwidth of different ultrabroadband OPAs. The three-stage OPAs yield few-cycle pulses covering different spectral regions (Vis, NIR, IR) with energies of 0.1-0.5 mJ. The NIR and IR pulses are coherently combined to two octaves (or more) of synthesized bandwidth corresponding to sub-cycle pulses. An active timing stabilization system based on in-line multi-phase meters and FPGA-based single-shot phase retrieval allows for stabilization and control of repeatable pulses over hours of operation, allowing even for demanding experimental applications such as attosecond streaking and hence suitable for attosecond-resolved experiments. A reduced basis set of CEP and RP's allows accessing almost the complete plethora of possible waveforms to be synthesized while keeping the system complexity low enough to allow reliable operation. A modular optical setup exhibiting superb timing stability and careful spatial overlap characterization allows for the next generation of attosecond pump-probe spectroscopic applications. In the next step, a third spectral channel will be fully integrated to yield an additional pulse in the visible spanning 520-700\,nm with 150\,$\upmu$J of pulse energy and 6\,fs in duration. This synthesis allows for pulses to be as short as 1.9\,fs in duration and with even more intricate non-sinusoidal waveform customization. Furthermore, additional degrees of freedom for waveform design arise. At this level, a genetic algorithm of adaptive waveform control might become necessary to help optimize the IAP characteristics, as plain waveform scanning would take multiple days due to the additional degrees of freedom.\\
The PWS proved wide-bandwidth pulse generation. The approach will be fused together with novel pump laser technologies in the next development step. Due to the multiple OPA-stages, a parallel laser amplifier scheme can be used to increase the pulse energy, average power, and repetition rate of the system.\\
Sub-cycle tailored waveforms capable of driving tunable isolated attosecond pulse generation over a broad energy range (up to the soft x-ray) will open up new possibilities for attosecond pump-probe experiments. This includes schemes where by utilizing sub-cycle waveforms to drive strong-field ionization will be possible to confine the pump-excitation to few hundreds of attosecond, enabling as-pump/as-probe resolution.

\section*{Acknowledgement}
We gratefully acknowledge support from Deutsches Elektronen-Synchrotron (DESY) of the Helmholtz Association, from the Cluster of Excellence ‘CUI: Advanced Imaging of Matter’ of the Deutsche Forschungsgemeinschaft (DFG)—EXC 2056—project ID 390715994, the priority programme ‘Quantum Dynamics in Tailored Intense Fields’ (QUTIF) of the DFG (SPP1840 SOLSTICE) and H2020 European Research Council (ERC) (FP7/2007-2013, ERC 609920-AXSIS).\\
G. M. Rossi would like to thank Dr. Audrius Zaukevicius for fruitful discussions.


\FloatBarrier
\newpage
\section*{References}
\bibliographystyle{iopart-num}
\bibliography{biblio}

\end{document}